\documentclass{article}

\PassOptionsToPackage{sort,numbers,compress}{natbib}

     \usepackage[final]{neurips_2024}

\usepackage[utf8]{inputenc} 
\usepackage[T1]{fontenc}    
\usepackage{hyperref}       
\usepackage{url}            
\usepackage{booktabs}       
\usepackage{amsfonts}       
\usepackage{nicefrac}       
\usepackage{microtype}      
\usepackage{xcolor}         

\usepackage[pdftex]{graphicx}
\usepackage{blindtext}
\usepackage{multirow}
\usepackage{amsmath}
\usepackage{amsthm, amssymb}
\usepackage{wrapfig}
\usepackage{pifont}
\usepackage{enumitem}

\theoremstyle{plain}

\newtheorem{proposition}{Proposition}

\title{Differentiable Modal Synthesis for Physical Modeling of Planar String Sound and Motion Simulation}

\author{%
  Jin Woo Lee
  \thanks{Music and Audio Research Group (MARG), Department of Intelligence and Information} \\
  Seoul National University \\
  \texttt{jinwlee@snu.ac.kr} \\
  \And
  Jaehyun Park
  \footnotemark[1] \\
  Seoul National University \\
  \texttt{lotussoh@snu.ac.kr} \\
  \AND
  Min Jun Choi
  \footnotemark[1] \\
  Seoul National University \\
  \texttt{choimj21@snu.ac.kr} \\
  \And
  Kyogu Lee
  \footnotemark[1]
  ~\thanks{Artificial Intelligence Institute of Seoul National University (AIIS)}
  ~\thanks{Interdisciplinary Program in Artificial Intelligence (IPAI)}
  \\ Seoul National University \\
  \texttt{kglee@snu.ac.kr} \\
}

\begin{document}

\maketitle

\begin{abstract}
While significant advancements have been made in music generation and differentiable sound synthesis within machine learning and computer audition, the simulation of instrument vibration guided by physical laws has been underexplored. To address this gap, we introduce a novel model for simulating the spatio-temporal motion of nonlinear strings, integrating modal synthesis and spectral modeling within a neural network framework. Our model leverages physical properties and fundamental frequencies as inputs, outputting string states across time and space that solve the partial differential equation characterizing the nonlinear string. Empirical evaluations demonstrate that the proposed architecture achieves superior accuracy in string motion simulation compared to existing baseline architectures. The code and demo are available online.
\footnote{\url{https://huggingface.co/spaces/szin94/dmsp}}
\end{abstract}

\section{Introduction}

The investigation of wave propagation along strings, encompassing both theoretical and experimental dimensions, has persisted for well over a century \cite{rayleigh1896theory,donkin1884acoustics}.  In the relentless pursuit of verisimilitude and expressive fidelity in simulating wave phenomena, numerous studies have been investigated to bridge the gap between theoretical underpinnings and empirical sound measurements \cite{fletcher1964normal}.  Advancements leveraging computational power to mimic the intricate physical processes in musical instruments have given rise to numerical sound synthesis, now a cornerstone in the field of virtual sound synthesis \cite{ruiz1970technique,hiller1971synthesizing}.  \citet{schwarz2007corpus} presents a systematic study of parametric and physical models for music audio synthesis.  Parametric models include signal models, such as spectral modeling synthesis \cite{serra1990spectral}. Physical models encompass techniques such as modal synthesis, digital waveguides \cite{smith2010physical,fettweis1986wave}, or finite-difference time-domain (FDTD) methods \cite{ruiz1970technique,bilbao2009numerical}.

Over recent years, the advancement of hardware acceleration for artificial intelligence has enabled the emergence of numerous techniques for neural audio synthesis \cite{hayes2024review}, including autoregressive generation \cite{wavenet}, adversarial training \cite{donahue2018adversarial} with phase coherence \cite{engel2018gansynth}, and approximated physical models \cite{valin2019lpcnet}.  The concept of differentiable digital signal processing (DDSP) was first introduced by \citet{engel2019ddsp}, aiming to incorporate a known signal model into neural networks to achieve a domain-appropriate inductive bias.  While the DDSP model can be considered a differentiable version of spectral modeling synthesis, a wide variety of works have explored the differentiable implementation of other audio signal processing methods, such as the subtractive method \cite{wang2019neural}, waveshaping \cite{hayes2021neural}, and frequency modulation \cite{caspe2022ddx7,Braun_DX7-JAX_2023}.  Subsequent research has demonstrated various applications of DDSP, including music performance synthesis \cite{wu2021midi}, speech synthesis and voice conversion \cite{choi2021neural,choi2022nansy}, and sound effect generation \cite{barahona2024noisebandnet}.  \citet{renault2022differentiable} extended DDSP to create a polyphonic synthesizer, explicitly modeling properties specific to piano strings, such as inharmonicity and detuning induced by string stiffness, based on a parametric model of these phenomena \cite{rigaud2011parametric}. This model efficiently synthesizes piano sound from MIDI input, achieving a high mean opinion score on naturalness.  However, it still shows room for improvement compared to sampling-based methods \cite{henningsson2011fluidsynth} and physical modeling methods \cite{bank2018model}.

Despite the growing recognition of DDSP as a promising sound synthesis methodology, its extension to physical modeling remains underexplored.
\citet{schlecht2022physical} have presented physical modeling using Fourier Neural Operator (FNO).
They train recurrent-type FNOs to learn state transitions from data spanning a few initial timesteps in the simulation and then test generalization to the subsequent long-range data.
Although they have demonstrated encouraging outcomes, their approach has room for improvement in that it is unconditional, meaning that it is challenging to generalize over dynamic scenarios (\textit{e.g.}, glissando, vibrato).
In the context of rigid-body contact sound synthesis \cite{james2006precomputed,wang2019kleinpat,o2002synthesizing}, a few studies investigated the efficacy of training neural networks supervised by finite-element method (FEM) solvers.
\citet{jin2020deep,jin2022neuralsound} propose a neural network that predicts contact sounds from voxelized objects, inspired by the modal technique that synthesizes sound using eigenvalues and eigenvectors.
\citet{diaz2023rigid} leverage a differentiable infinite impulse response filter to synthesize contact sounds from rasterized occupancy grids in an end-to-end manner.
These methods can interactively synthesize sounds for various contact conditions and materials with notable efficiency, circumventing the need for an offline optimization process typical of modal techniques.
However, these methods, which resort to the FEM solver, are vulnerable in modeling the dynamic behavior or in simulating the motion of the object.

\begin{figure}
    \centering
    \includegraphics[width=\linewidth]{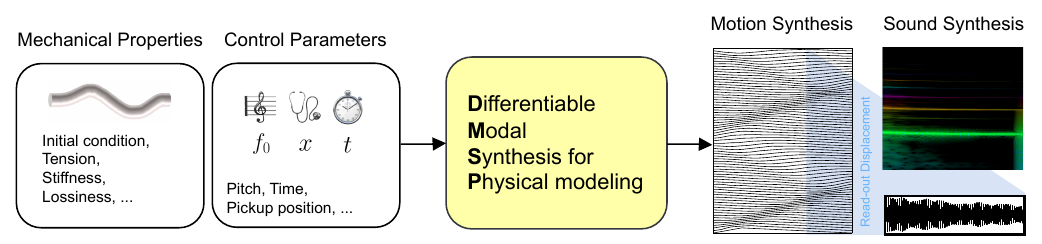}
    \caption{\textbf{System overview}.
         The DMSP model encodes the physical properties of a string
         (e.g., tension, stiffness, damping, and initial conditions)
         to estimate the displacement of the string plucked at pitch \(f_0\)
         at a given time \(t \in [0, \infty)\) and position \(x \in \Omega\).
         By concatenating the DMSP outputs over the domain \((x, t) \in \Omega \times [0, \infty)\),
         the simulated motion of the string can be visualized.
         Reading the outputs at a particular position \(x\)
         allows hearing the synthesized string sound,
         akin to listening with a stethoscope at the pickup position.
    }
    \label{fig: system overview}
\end{figure}

In this regard, we propose a novel model for simulating the spatio-temporal motion of nonlinear strings, integrating modal synthesis and spectral modeling within a neural network framework.  The proposed model leverages physical properties and fundamental frequencies as inputs, outputting string states across time and space that solve the partial differential equation (PDE) characterizing the nonlinear string.  Empirical evaluations demonstrate that the proposed architecture achieves superior accuracy in string motion simulation compared to the baseline architectures.
The main contributions are as follows:
\begin{itemize}[leftmargin=5mm]
    \item We present differentiable modal synthesis for physical modeling (DMSP)
          that simulates dynamic nonlinear string motion by synthesizing sound
          using the physical properties of the string.
    \item To the best of our knowledge, this is the first differentiable approach
          that can synthesize the motion and the sound of musical strings
          with a dynamic control over the pitch and the material properties.
    \item We provide an extensive empirical evaluation demonstrating the
          importance of modal decomposition and the proper choice of loss function.
\end{itemize}

\section{Background}

\subsection{Physical Modeling of Musical String Instrument}
\label{ssec: physical modeling}
\textbf{Linear Damped Stiff String.}
The string model discussed in this paper is a damped nonlinear stiff string.
To introduce the nonlinear string, we first formulate the governing equations
for the damped linear stiff string system and derive the corresponding modal solution.
\begin{equation}\label{eqn:linear-wave}
    \partial_{tt} u = \gamma^2 \partial_{xx} u - \kappa^2 \partial_{xxxx} u - 2\sigma_0 \partial_{t} u
\end{equation}
The linear string of its length \(L\), vibrating with wave speed \(\gamma\),
stiffness \(\kappa\), and frequency-independent damping factor \(\sigma_0\),
is described by \autoref{eqn:linear-wave}.
Given the initial conditions (IC) for \(x\in\Omega=[-L/2,+L/2]\) as
\(u(x,0) = u_0(x)\) and \(\partial_t u(x,0) = 0\),
and appropriate boundary conditions (BC), the corresponding solution \(u(x,t)\)
represents the motion of a damped linear stiff string.
Particularly, for a clamped boundary condition,
\textit{i.e.}, \(u(\pm L/2, t) = \partial_x u(\pm L/2, t) = 0\) for all \(t\in[0,\infty)\),
a modal solution can be obtained as follows.
\begin{subequations}\label{eqn: linear analytic}
    \begin{align}
        u(x,t) &= \sum_{n=1}^\infty X_n(x) T_n(t) \label{eqn: analytic-u}\\
        X_n(x) &= c_1\left(\sin\mu_n x - \frac{\sin\mu_n L/2}{\sinh\nu_n L/2}\sinh\nu_n x\right) + c_2\left(\cos\mu_n x - \frac{\cos\mu_n L/2}{\cosh\nu_n L/2}\cosh\nu_n x\right) \label{eqn: analytic-X}  \\
        T_n(t) &= e^{-\sigma_0 t} \cos\underbrace{\sqrt{\mu_n^4\kappa^2 + \mu_n^2 \gamma^2 - \sigma_0^2}}_{\omega_n} t \label{eqn: analytic-T}
    \end{align}
\end{subequations}
The derivation of \autoref{eqn: linear analytic} can be found in
\autoref{appendix: linear string vibration}.
The allowed values of \(\mu_n\) and \(\nu_n\) are determined by the boundary conditions,
while the coefficients \(c_1\) and \(c_2\) are determined using the
initial condition \(\sum_{n=1}^\infty X_n(x) = u_0(x)\).
Determining these values typically requires an offline numerical solving process,
where the obtained values can be stored in memory for real-time computation of \(u(x,t)\).
The stiffness modeled by the 4\textsuperscript{th} order derivative induces a hyperbolic solution, resulting in non-integer multiples of the mode frequencies \(\omega_n\), which leads to physical inharmonicity. The damping factor \(\sigma_0\) causes an exponential decay in the temporal amplitude.

\setlength{\intextsep}{9pt}%
\setlength{\columnsep}{9pt}%
\begin{wrapfigure}{r}{0.23\linewidth}
  \centering
  \vspace{-8mm}
  \includegraphics[width=\linewidth]{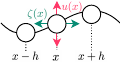}
  \caption{The planar string system.}
\end{wrapfigure}
\textbf{Nonlinear Damped Stiff String.}
The generalization of the linear wave \autoref{eqn:linear-wave} to nonlinear string vibrations is first introduced by Kirchhoff \cite{kirchhoff1897vorlesungen} and Carrier \cite{carrier1945non}.
The Kirchhoff--Carrier system models elastic strings in two dimensions,
and when extended so that transverse and longitudinal motions are coupled,
phantom partials can be exhibited, resulting in a richer timbre \cite{bilbao2023models}.
A model of such planar string vibration is as follows \cite{morse1986theoretical,anand1969large}.
\begin{subequations}\label{eqn: wave-nonlinear}
    \begin{align}
        \partial_{tt} u &= \gamma^2 \partial_{xx} u - \gamma^2\frac{\alpha^2-1}{2}\partial_x\left(q^3+2pq\right) - \kappa^2\partial_{xxxx} u - 2\sigma_0\partial_t u + 2\sigma_1\partial_t\partial_{xx} u                    \label{eqn: wave-u-nonlinear}\\
        \partial_{tt}\zeta &= \gamma^2\alpha^2 \partial_{xx}\zeta - \gamma^2\frac{\alpha^2-1}{2}\partial_x\left(q^2\right)  - 2\sigma_0\partial_t \zeta + 2\sigma_1\partial_t\partial_{xx} \zeta        \label{eqn: wave-z-nonlinear}
    \end{align}
\end{subequations}
Here, \(u(x,t)\) and \(\zeta(x,t)\) represent the transverse and longitudinal displacements of a string, respectively, for all \((x,t) \in \Omega \times [0,\infty)\).
\(q = \partial_x u\) and \(p = \partial_x \zeta\) serves as the auxiliary coupling system, as \(\partial_t q = \partial_x \partial_t u\) and \(\partial_t p = \partial_x \partial_t \zeta\) \cite{bilbao2005conservative}.
A more detailed background on the derivation of \autoref{eqn: wave-nonlinear} can be found in \autoref{appendix: planar string vibration}.
The boundary condition, which may vary depending on the string being modeled, is chosen to be that of the clamped boundary condition:
\begin{equation}\label{eqn: clamped-bc}
    u(x,t) = \partial_x u(x,t) = 0, \quad \forall (x,t) \in \partial \Omega \times [0,\infty).
\end{equation}
Given an initial condition \(u_0 := u(x,0)\) defined on \(x \in \Omega\), the solution \(u(x,t)\) associated with the condition of \autoref{eqn: clamped-bc} simulates the motion of the string for physical modeling and sound synthesis of string instruments.
Due to the coupling between \autoref{eqn: wave-u-nonlinear} and \autoref{eqn: wave-z-nonlinear},
the obtained solution exhibits features found in elastic strings, such as pitch glide and phantom partials.
These features become more pronounced for larger displacements and are difficult to approach separately as in the linear case.
The solution can be approximated through various physical modeling techniques such as finite-difference time-domain \cite{bilbao2004energy}, digital waveguides \cite{maestre2017joint}, or functional transformation method \cite{trautmann2004multirate}.

\textbf{Finite-difference Time-domain.}
One straightforward approach to tackling nonlinear PDEs such as \autoref{eqn: wave-nonlinear} would be employing finite difference approximation.  This method, commonly known as finite-difference time-domain (FDTD), has a long and distinguished history and is widely accepted in fields such as fluid dynamics \cite{moin1998direct} and electromagnetics \cite{taflove2013advances}.  FDTD is particularly effective in solving nonlinear, multidimensional, and dynamic systems.  The extensive literature on its applications encompasses a diverse range of domains, including musical acoustics \cite{bilbao2009numerical}.  A recent contribution by \citet{lee2024string} introduces StringFDTD-Torch, an FDTD simulator tailored for modeling planar damped stiff strings akin to \autoref{eqn: wave-nonlinear}.  Leveraging PyTorch C++ extension, StringFDTD-Torch facilitates FDTD computations on both CPUs and GPUs.  However, its current iteration lacks support for gradient backpropagation through the FDTD module, leaving room for enhancement, particularly in optimizing the gradient computation process, which is hindered by the substantial number of temporal recursions involved (evident by the large $N_t$ in $\mathcal{O}(N_x N_t)$ of \autoref{tab: comparison}).

\textbf{Modal Synthesis.}
As a more efficient approach to solve the nonlinear wave equations, the modal synthesis \cite{adrien1991missing,adrien1985physical,morrison1993mosaic} decomposes the complex dynamics into contributions from a set of modes, whose spatial bases are eigenfunctions of the pertinent problem. Each mode exhibits distinct oscillations at complex frequencies, contingent upon the boundary conditions.  For problems with real-valued parameters, these complex frequencies occur in conjugate pairs, and the "mode" is thus defined as the pair of such eigenfunctions and frequencies \cite{bilbao2009numerical}.  Modal synthesis involves two primary steps.  Initially, in an offline phase (also labeled as `Pre-computation' in \autoref{tab: comparison}), modal shapes and frequencies are discerned from the PDE system, considering both boundary and initial conditions.  This information is encapsulated in what is known as a shape matrix.  Subsequently, the solution is derived by combining the modal functions, each progressing at its natural frequency.  Pre-computation requires $\mathcal{O}(N_m)$ of recursions because $N_m$ shape matrices need to be computed, but a typical $N_m$ is typically hundreds to thousands of times less for $N_t$.  Modal synthesis after this off-line process is very efficient as it allows us to obtain a solution for a given $x$ and $t$ without any recursion, but it is clear that the range of solutions that can be covered is bounded in that it relies on the ansatz $u(x,t)=X(x)T(t)$ for the separation of variables.

\subsection{Differentiable Digital Signal Processing}
In the field of neural networks, numerous audio researchers have been engaged in the development of techniques that leverage the ease of automatic gradient backpropagation in neural networks for the purpose of audio parameter estimation.  To address the challenge of estimating the latent parameters of a sound, some approaches implement the synthesis part as-is using automatic differentiation package \cite{lee2024string,Braun_DX7-JAX_2023} so that the gradient can back-propagate through it to update the parameters directly, while the majority of approaches train neural networks to estimate the parameters in an auto-encoder framework \cite{engel2019ddsp,lee2022differentiable,choi2021neural}.  As one of the most seminal studies of the latter approach, DDSP is widely used to efficiently synthesize nonlinear and dynamic sounds.  Based on the spectral modeling synthesis framework, the time-domain signal is modeled via short-time Fourier transforms (STFTs) divided into deterministic (harmonic) and stochastic (noisy) parts to synthesize the sound.  As the causality of STFT frames is modeled through gated recurrent units (GRUs), DDSP requires as many recursions as $N_r$, the number of frames. The harmonics of a DDSP are given by an integer multiple of the fundamental frequency ($f_0$), and the noise is synthesized from filtered noise.  These DDSPs, while still a remarkable advancement, have strong structural constraints on the deterministic part to capture enough perceptually rich tones such as inharmonicity due to stiffness or phantom partials due to nonlinearity.  A study by \citet{renault2022differentiable} also points this out, and uses the parametric model \cite{rigaud2011parametric} for inharmonicity and detune, but there is room for improvement as it is an approximation model that relies on instrument-specific modifiers rather than reflecting stiffness physics.

\begin{table}
  \vspace{-9mm}
  \centering
  \caption{
      Comparison between methods. Computational complexity refers to the inference scenario.
  }
  \label{tab: comparison}
  \begin{tabular}{lccccl}
    \toprule
    \multirow{2}[2]{*}{\textbf{Method}} & \multicolumn{3}{c}{\textbf{System taxonomies}} & \multicolumn{2}{c}{\textbf{Computational complexity}} \\
    \cmidrule(r){2-4} \cmidrule(r){5-6}
    & Physical & Nonlinear & Differentiable
	& Pre-computation & Synthesis \\
    \midrule
    Modal  & \ding{51} & \ding{55} & \ding{55} & $\mathcal{O}(N_m)$ & $\mathcal{O}(1)$ \\
    FDTD   & \ding{51} & \ding{51} & \ding{55} & N/A  & $\mathcal{O}(N_x N_t)$  \\
    DDSP   & \ding{55} & \ding{51} & \ding{51} & N/A  & $\mathcal{O}(N_r)$  \\
    \midrule
    \textbf{DMSP-Hybrid} & \ding{51} & \ding{51} & \ding{51} &  $\mathcal{O}(N_m)$ & $\mathcal{O}(1)$  \\
    \textbf{DMSP}   & \ding{51} & \ding{51} & \ding{51} & N/A & $\mathcal{O}(1)$  \\
    \bottomrule
  \end{tabular}
\end{table}

\section{Differentiable Modal Synthesis for Physical Modeling (DMSP)}
This section introduces a novel differentiable nonlinear string sound synthesizer.  \autoref{tab: comparison} provides a summary of the methods discussed.  While modal synthesis stands out for its efficiency, it is solely applicable to linear models and necessitates pre-computation to determine the number of modes denoted as \(N_m\).  On the other hand, FDTD computes highly nonlinear and dynamic solutions but demands a substantial computational load due to iterative updates across both temporal (\(N_t\)) and spatial (\(N_x\)) samples.  In contrast, differentiable audio processing methods offer efficient nonlinear sound synthesis, typically leveraging a smaller number of frames (\(N_r\)) compared to the total time samples (\(N_t\)).  However, they often lack physical controllability.  In this regard, we propose DMSP, which approximates the solution of \autoref{eqn: wave-nonlinear} efficiently by leveraging \autoref{eqn: linear analytic} in the form of neural networks. \autoref{fig: system overview} provides a visual depiction of the DMSP.

\begin{figure}
	\begin{minipage}[b]{0.76\linewidth}
        \centering
        \centerline{\includegraphics[width=\linewidth]{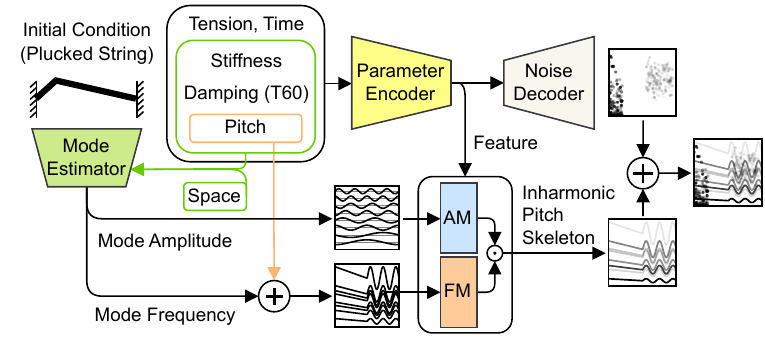}}
        \centerline{\footnotesize (a) Network architecture of DMSP}
    \end{minipage}
	\begin{minipage}[b]{0.23\linewidth}
        \centering
        \centerline{\includegraphics[width=\linewidth]{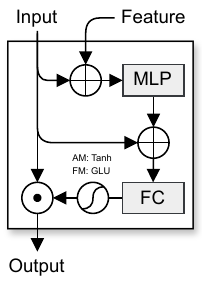}}
        \centerline{\footnotesize (b) AM/FM Block}
    \end{minipage}
    \caption{\textbf{Network architecture}.
DMSP synthesizes a pitch skeleton with an inharmonic structure,
drawing upon overtones derived from the modes of the string.
The modes can either be derived directly using the modal decomposition (DMSP-Hybrid, the hybrid of DMSP and Modal),
or using the neural network trained to estimate the modes (DMSP, the fully-neural-network method).
Yet, relying solely on modal frequencies and corresponding shape functions
delineates a linear solution, which falls short of capturing the nuances of nonlinear string motion.
To address this, DMSP introduces FM and AM blocks to modulate the
modes of the linear solution. This modulation process enables DMSP to
estimate the pitch skeleton of the nonlinear solution.
Consequently, the output waveform is synthesized through the spectral modeling pipeline,
incorporating both (in)harmonic components and the filtered noise.
    }
	\label{fig: network}
\end{figure}

\subsection{Problem Statement}
The objective of this study is to establish a mapping from the parameter space
to the solution space, utilizing a finite set of observations comprising
parameter-solution pairs from this mapping. We delineate this objective as follows:
Consider the partial differential equation depicted in \autoref{eqn: wave-nonlinear},
applicable for \((x,t)\) within \(\Omega \times [0,\infty)\),
with clamped boundary conditions as specified in
\autoref{eqn: clamped-bc},
where \(\Omega\) represents a bounded domain in \(\mathbb{R}^{N_x}\).
We assume that the solution \(u:\Omega \times [0,\infty) \to \mathbb{R}\)
resides within the Banach space \(\mathcal{U}\).
For a given PDE parameter \(\rho \in \mathcal{P}\) and initial condition \(u_0 \in \mathcal{U}\),
let \(\mathcal{S}:\mathcal{P} \to \mathcal{U}\) denote a nonlinear map,
specifically, the FDTD numerical solver tailored to the context of this study.
Assume that we are provided with observations \(\{\rho^{(i)}, u^{(i)}\}_{i=1}^{N}\), where \(\rho^{(i)}\) comprises independent and identically distributed (i.i.d.) samples
drawn from a probability measure supported on \(\mathcal{P}\), and \(u^{(i)} = \mathcal{S}(\rho^{(i)})\) potentially contains noise.
Our goal is to construct an approximation of \(\mathcal{S}\) denoted as \(\mathcal{S}_\theta:\mathcal{P} \to \mathcal{U}\),
and select parameters \(\theta^* \in \mathbb{R}^{N_\theta}\) such that \(\mathcal{S}_{\theta^*} \approx \mathcal{S}\).
Leveraging \(\mathcal{S}_\theta\), one can compute the solution \(\hat{u} = \mathcal{S}_\theta(\rho)\) corresponding to
a new parameter \(\rho \in \mathcal{P}\). By specifying values for \(x\) and \(t\), one can then either synthesize the
sound of the string picked-up (also referred to as read-out) at a specific location \(x_0\) as \(\hat{u}(x_0,t)\),
or simulate the motion of the string by concatenating \(\hat{u}(x,t)\) across all \(x \in \Omega\).
In practice, \(\Omega\) and \([0,\infty)\)
are bounded and discretized to form an evenly distributed spatio-temporal grid as
\(\mathbb{R}^{N_x}\times\mathbb{R}^{N_t}\subset\Omega\times[0,\infty)\),
where the spatial grid samples are uniformly distributed in space
according to the length of the equally spaced intervals \(L/N_x\)
and the temporal grid samples uniformly distributed according
to the frequency of a fixed audio sampling rate.

\subsection{Network Architecture}

\textbf{Parameter Encoder.}
To effectively capture material features inherent in the PDE parameter values, the parameter encoder leverages a random Fourier feature (RFF) layer \cite{rahimi2007random,tancik2020fourier}. Given T60 frequencies \(f_{\mathrm{T60}}^{(i)}\) and their corresponding times \(t_{\mathrm{T60}}^{(i)}\) for \(i=1,2\), the frequency-dependent damping coefficients \(\sigma_0\) (and \(\sigma_1\), if applicable) are derived using \autoref{eqn: t60}. The frequency-independent damping factor \(\exp(-\sigma_0 t)\) is computed explicitly multiplied by the mode amplitudes. All PDE parameters \(\rho=\left\{\kappa,\alpha,\sigma_0,\sigma_1\right\}\in\mathbb{R}^4\subset\mathcal{P}\) are encoded into a feature vector \(h\in\mathbb{R}^{4 \times d}\) with a Fourier embedding size of \(d=256\).

\textbf{AM and FM Blocks.}
As illustrated in \autoref{fig: network}, DMSP employs amplitude modulation (AM) and frequency modulation (FM) modules
\footnote{
It is worth noting that there are some subtle differences from the AM/FM techniques commonly used in sound synthesis and the usage of the terminology in this paper.
Relying on terminology commonly used in AM/FM technique, AM/FM in this paper can be described analogically as a carrier sinusoid whose center frequency and amplitude are determined by modal decomposition and varied by a non-sinusoidal modulator.
While typical AM/FM synthesis also uses non-sinusoidal modulators, the intent of this paper is different in that it borrows these for the purpose of limiting the undesired periodicity in those modulators.
}
to modulate the mode frequencies and amplitudes of the linear solution, as depicted in \autoref{eqn: linear analytic}, to synthesize the solution of the nonlinear wave described in \autoref{eqn: wave-nonlinear}.
We utilize two multilayer perceptrons (MLPs) for the modulation layers.
Although a simpler and perhaps more conventional choice of architecture would be a GRU,
to the decoder architecture of DDSP \cite{engel2019ddsp}, we choose MLPs for their slightly better empirical results.
It's noteworthy that DDSP decodes the sinusoidal frequency envelope with fixed frequency values. In contrast, DMSP decodes both the envelope and the frequency values independently, employing two distinct MLP blocks, namely AM and FM.

\textbf{Mode Estimator.}
As detailed in \autoref{ssec: physical modeling}, determining allowed values for mode frequencies and amplitudes, corresponding to specific initial and boundary conditions of the string, typically involves a root-finding process conducted offline. While these numerical solvers offer high accuracy up to a specified iterative threshold, they necessitate pre-computation, as illustrated in \autoref{tab: comparison}.
The mode estimator module within DMSP estimates the modes from the initial condition using an MLP. The initial condition is parameterized by a pluck position \(p_x\) and its peak amplitude \(p_a\). Subsequently, the physical properties \(\kappa\), \(\gamma\), \(\sigma_0\), and \(\sigma_1\) are encoded using a random Fourier feature (RFF) layer. The mode frequencies and amplitudes are then estimated by the MLP, followed by the application of suitable scaling activations.
It's pertinent to note that the mode estimator operates independently and is trained separately from the other modules. During training, the ground truth modes (computed using the modal decomposition method) are fed into the AM and FM blocks, ensuring accurate synthesis while training the synthesis part of the model.

\subsection{Loss Function}
We employ a combination of four loss terms:
(1) waveform \(\ell_1\) loss (\(\mathcal{L}_1\)) that measures the \(L_1\) discrepancy between the synthesized waveform and the ground truth waveform,
(2) Multi-scale spectral (MSS) loss that captures spectral differences across multiple scales, ensuring fidelity in spectral representation,
(3) Pitch loss (\(\mathcal{L}_{f_0}\)) that penalizes deviations from the ground truth fundamental frequency (\(f_0\)), and
(4) Mode loss (\(\mathcal{L}_{m}\)), which measures the \(L_1\) distance of the mode frequency and mode amplitude from the output of the mode estimator (if applicable) and the mode frequency and mode amplitude obtained via modal decomposition.
MSS loss has been adopted as a metric for reconstruction in most neural net-based synthesis techniques \cite{engel2019ddsp,wu2021midi,hayes2024review}.
Concerning that measuring MSS with magnitudes is not phase-sensitive by definition, we employ the \(\mathcal{L}_1\) loss in the waveform to train the causality in spatio-temporal wave propagations.
Challenges in optimizing the frequency parameters of sinusoidal oscillators via gradient descent over the spectral loss functions, due to the non-convex nature of the optimization problem, have been highlighted in various studies \cite{turian2020m,hayes2023sinusoidal}.
Damped sinusoids offer a workaround for the issue of non-convexity concerning frequency parameters \cite{hayes2023sinusoidal}, or alternative metrics are proposed to mitigate the risk of falling into bad local minima \cite{torres2024unsupervised,schwar2023multi}.
We adopt a parameter regression joint training strategy, akin to the pre-training phase of the work by \citet{engel2020self}.
Among the output mode frequencies, we train the model to match one mode frequency component (denoted by $\hat{f}_0$) to match the fundamental frequency of the target FDTD-simulated audio (\(f_0\)) annotated using CREPE \cite{kim2018crepe} as \(\mathcal{L}_{f_0}=\|\hat{f}_0-f_0\|_1\).

\section{Experiments}

\subsection{Experimental Setup}\label{ssec: experimental setup}

\textbf{Dataset.}
We use the StringFDTD-Torch \cite{lee2024string}, the
open-source nonlinear string simulator,
to compute the solution of the \autoref{eqn: wave-nonlinear} in
a temporal sampling rate of 48 kHz and a spatial sampling rate.
The solution is upsampled to a spatial resolution of 256 using a
bivariate spline approximation over a rectangular spatio-temporal mesh
upto the 5\textsuperscript{th} order degree. 
We simulate 10263 different strings by randomly augmenting
the material properties, \textit{e.g.}, \(\kappa\), \(\alpha\),
\(\sigma_0\), and  \(\sigma_1\), with various plucking profiles \(u_0\).
The simulation results in a total amount of 729.8 hours of wave files
that corresponds to the 1-sec string sounds picked up at 256 different positions
for each string.
For the test data, 715 strings are newly synthesized with the parameters sampled in i.i.d.
The test data consists of 336 linear (\(\alpha=1\)) and 379 nonlinear (\(\alpha>1\)) strings,
each of which has a 256 spatial grid size, resulting in approximately 50 hours of wave files.
\autoref{tab: params} specifies the range of the sampled PDE parameters.

\textbf{Baselines.}
\autoref{tab: comparison baselines} compares the major differences between the baselines and the proposed models.
As the first attempt to tackle the task of neural audio synthesis for dynamic physical properties, we compare our proposed model to three other models.
We consider three baselines, namely: Modal and DDSPish, and DDSPish-{\scriptsize XFM}.
Modal synthesis is the linear wave solution as in \autoref{eqn:linear-wave},
where the modal frequencies and the shape functions are pre-computed.
DDSPish is a neural network based on the harmonic plus noise model, similar to DDSP.
Yet, this -ish suffix emphasizes that this model is
different from the DDSP model \cite{engel2019ddsp}.
DDSPish does not have a reverb module but instead adds frequency modulation,
and most notably it has a parameter encoder that allows
generating sounds from physical parameters.
Please see \autoref{appendix: baseline} for more details of the baselines.

\textbf{Evaluation Metrics.}
We report results on three metrics,
signal-distortion-ratio (SDR),
scale-invariant signal-distortion-ratio (SI-SDR) \cite{le2019sdr},
multi-scale spectral (MSS) distance,
and the pitch difference in Hz.
As with the problem statement, we consider the FDTD-simulated results as the ground truth (GT).
Both SDR and SI-SDR estimate the distortions on a spatiotemporal grid and are very strict about scoring out-of-phase cases, by directly comparing the the estimated displacements to the FDTD results without any specific transformations or interpolations.
Given that the magnitudes of the estimated displacements typically distribute within the small range (approximately between $\pm 0.02$; depends on the pluck amplitude $p_a$, see \autoref{tab: params}), we compare both with scale normalization (SI-SDR) and without (SDR).
The MSS metric for the evaluation is computed using the short-time Fourier transformation (STFT) with three scales of Fast Fourier Transform (FFT) points:
1024, 512, and 256, with a window length equal to the FFT point for each, and a hop length equal to a quarter of that.
For each scale, the STFT magnitude is compared on both linear and log scales, weighted by 2.0 and 0.5, respectively.
This choice of weighting is to help scale the loss computation with MSS and should not have a significant impact on performance, but is important to note for absolute comparisons of MSS scores.
The Pitch metric is computed as the $\ell^1$ norm of the difference between $\hat{f}_0$ and $f_0$.

\begin{table}
  \vspace{-9mm}
  \centering
  \caption{
      Comparison between the baselines and the proposed models.
  }
  \label{tab: comparison baselines}
  \begin{tabular}{l c cc cccc}
    \toprule
    \multirow{2}[2]{*}{\textbf{Method}} & \multirow{2}[2]{*}{\textbf{Center frequency}} & \multicolumn{2}{c}{\textbf{Network architecture}} & \multicolumn{4}{c}{\textbf{Training configuration}} \\
    \cmidrule(r){3-4} \cmidrule(r){5-8}
    & & AM block & FM block
	& $\mathcal{L}_1$ & MSS & $\mathcal{L}_{f_0}$ & $\mathcal{L}_m$ \\
    \midrule
    Modal                       & Mode frequency    & \ding{55} & \ding{55} &       N/A &       N/A &       N/A &       N/A \\
    DDSPish-{\scriptsize XFM}   & Integer multiples & \ding{51} & \ding{55} & \ding{51} & \ding{51} & \ding{55} & \ding{55} \\
    DDSPish                     & Integer multiples & \ding{51} & \ding{51} & \ding{51} & \ding{51} & \ding{55} & \ding{55} \\
    \midrule                                                                                       
    \textbf{DMSP-Hybrid}        & Mode frequency    & \ding{51} & \ding{51} & \ding{51} & \ding{51} & \ding{51} & \ding{55} \\
    \textbf{DMSP}               & Learnt estimates  & \ding{51} & \ding{51} & \ding{51} & \ding{51} & \ding{51} & \ding{51} \\
    \bottomrule
  \end{tabular}
\end{table}

\subsection{Results}

\begin{table}
  \caption{Synthesis Results}
  \label{tab: synthesis}
  \centering
  \begin{tabular}{l rrrr rrrr}
    \toprule
    \multirow{3}[2]{*}{\textbf{Model}} & \multicolumn{4}{c}{\textbf{Linear} ($\alpha=1$)} & \multicolumn{4}{c}{\textbf{Nonlinear} ($\alpha>1$)} \\
    \cmidrule(r){2-5} \cmidrule(r){6-9}
    & \textbf{SI-SDR} & \textbf{SDR}  & \textbf{MSS} & \textbf{Pitch} 
    & \textbf{SI-SDR} & \textbf{SDR}  & \textbf{MSS} & \textbf{Pitch} \\
    & (dB, $\uparrow$)  & (dB, $\uparrow$)	 & (dB, $\downarrow$) & (Hz, $\downarrow$) 
    & (dB, $\uparrow$)  & (dB, $\uparrow$)	 & (dB, $\downarrow$) & (Hz, $\downarrow$) \\
    \midrule
    Modal
        & \underline{--3.191}	 & \underline{0.681}	 & 18.449	& \underline{\textbf{0.420}}
        & --16.611	 & --1.900	 & 17.254	& 2.316  \\
    DDSPish
        & --39.478	 & --2.598	 & \underline{11.047}	 & 5.518
        & --25.951	 & --2.102	 & 9.745	 & 3.306  \\
    DDSPish-{\scriptsize XFM}
        & --46.609	 & --2.257	 & \underline{\textbf{10.911}}	 & 11.304 
        & --46.858	 & --2.272	 & 10.299	 & 14.013 \\
    \midrule
    \textbf{DMSP-Hybrid}
       	& \underline{\textbf{--2.844}}   & \underline{\textbf{1.496}}  & 12.525    & \underline{0.792}
        & \underline{\textbf{15.670}}    & \underline{\textbf{16.455}} & \underline{\textbf{4.772}}     & \underline{1.027}  \\
    \textbf{DMSP}
        & --22.298   & --2.000   & 12.504    & 1.717
        & \underline{--10.315}   &   \underline{0.221}   &  \underline{5.656}    & \underline{\textbf{1.437}} \\
    \bottomrule
  \end{tabular}
\end{table}

\textbf{Differentiable Sound Synthesis.}
The efficacy of DMSP is studied as shown in \autoref{tab: synthesis}.
First, the modal synthesis method calculates a linear solution for a damped stiff string,
and the mode for the linear test set is calculated offline.
The discrepancy between the outcomes of modal synthesis and FDTD in the linear case
can be attributed to the absence of frequency-dependent damping in \autoref{eqn:linear-wave}.
In other words, the frequency information serves as an oracle for the Modal, as evidenced by its lowest pitch score,
but the decay of amplitude over time is not sufficiently modeled, which is where the remaining models demonstrate superior performance.
Considering the DDSPish models, the MSS is approximately 1.6 dB ahead of the DMSPs, but the difference can reach up to 44 dB in the case of the SI-SDR.
It is worth noting that $\alpha$ is uniformly sampled within the range $(1,25)$ for the training data, falling into the category of the nonlinear strings.
Information about the ranges for sampling PDE parameters is in \autoref{tab: params}.
In the case of the nonlinear test set, the difference between the modal solution and the FDTD solution becomes larger.
On the other hand, DDSPish models, which are based on spectral modeling synthesis but learned without using any modal information, show the lowest performance.
The superiority of DMSP is even more pronounced in the nonlinear case.
While Modal, which can only cover the solution to the linear case, shows attenuated performance,
the DMSPs demonstrate the best performance in all metrics.
In particular, DMSP-Hybrid, which precomputes the mode frequency like Modal, performs FM and AM for the nonlinear solution,
showing an SI-SDR improvement of nearly 31 dB and an MSS improvement of nearly 13 dB over Modal.
DMSP, which estimates the mode as a neural network without the pre-computation step,
also outperforms Modal on all metrics for the nonlinear case.
The primary reason for the performance discrepancy between the nonlinear case and the linear case for DMSP is that the training data is rarely precisely equal to $1$ in the $\alpha$ distribution when it is sampled.

\begin{figure}
	\begin{minipage}[b]{0.24\linewidth}
        \centering
        \centerline{\includegraphics[width=\linewidth]{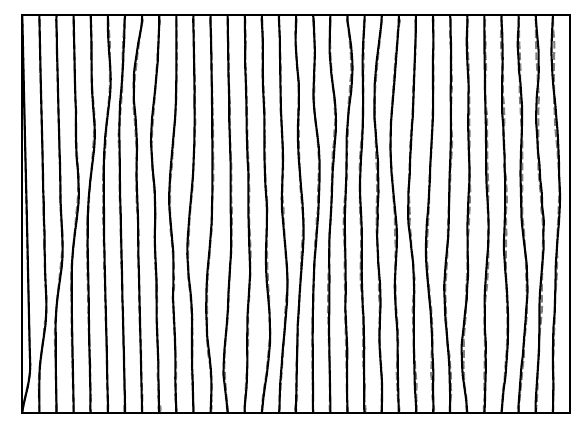}}
        \centerline{\footnotesize (a) $p_x=0.1$}
    \end{minipage}
	\begin{minipage}[b]{0.24\linewidth}
        \centering
        \centerline{\includegraphics[width=\linewidth]{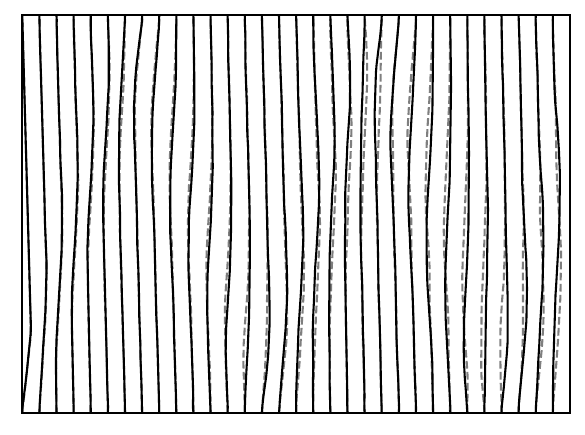}}
        \centerline{\footnotesize (b) $p_x=0.2$}
    \end{minipage}
	\begin{minipage}[b]{0.24\linewidth}
        \centering
        \centerline{\includegraphics[width=\linewidth]{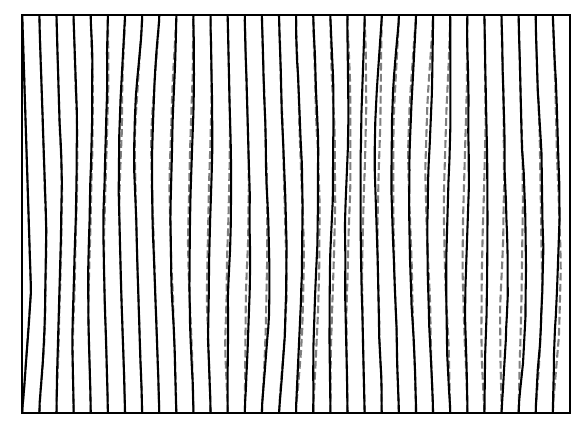}}
        \centerline{\footnotesize (c) $p_x=0.3$}
    \end{minipage}
	\begin{minipage}[b]{0.24\linewidth}
        \centering
        \centerline{\includegraphics[width=\linewidth]{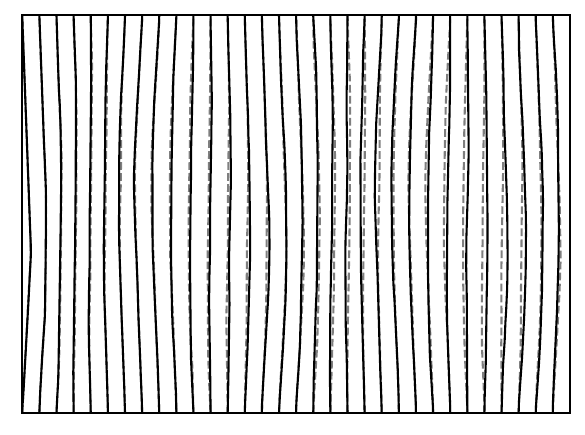}}
        \centerline{\footnotesize (d) $p_x=0.4$}
    \end{minipage}
    \caption{
        Visualization of the string displacement over time (horizontal) and space (vertical).
        For different initial conditions,
        the results synthesized by DMSP are shown as solid black lines and
        those simulated by FDTD as dashed gray lines.
    }
	\label{fig: px}
\end{figure}

\setlength{\intextsep}{9pt}%
\setlength{\columnsep}{9pt}%
\begin{wrapfigure}{r}{0.35\linewidth}
  \centering
  \vspace{-5mm}
  \includegraphics[width=\linewidth]{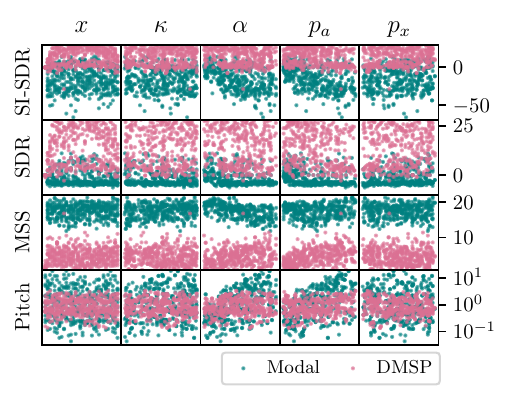}
  \vspace{-5mm}
  \caption{Objective scores over the change of physical parameters.}
  \vspace{-2mm}
  \label{fig: scatter}
\end{wrapfigure}
\textbf{Controllable Physical Simulation.}
The quantitative scores for various physical condition parameters are visualized in \autoref{fig: scatter}.
Trends show how the results of Modal synthesis and DMSP vary for different pickup positions ($x$), stiffness ($\kappa$), tension ($\alpha$), pluck amplitude ($p_a$), and pluck position ($p_x$).
Of these, $\alpha$ and $p_a$ in particular are known to increase the nonlinearity of the string as they increase in magnitude, which can be seen by the lower Modal synthesis scores. For DMSP, we see an overall improvement in the score, with a lower propensity for nonlinearity.
\autoref{fig: px} depicts the simulated state of the string as the pluck position in the initial condition is varied.
The results synthesized by DMSP can reconstruct a very accurate initial condition, similar to the results simulated by FDTD.
The vibration propagating through time along the string exhibits a distinct behavior contingent upon the initial condition.
FDTD employs a recursive calculation of displacement, necessitating some iterations equal to the number of samples at the audio sampling rate.
In contrast, DMSP is capable of obtaining the desired displacement in both time and space simultaneously.

\begin{figure}
	\begin{minipage}[b]{0.16\linewidth}
        \centering
        \centerline{\includegraphics[width=\linewidth]{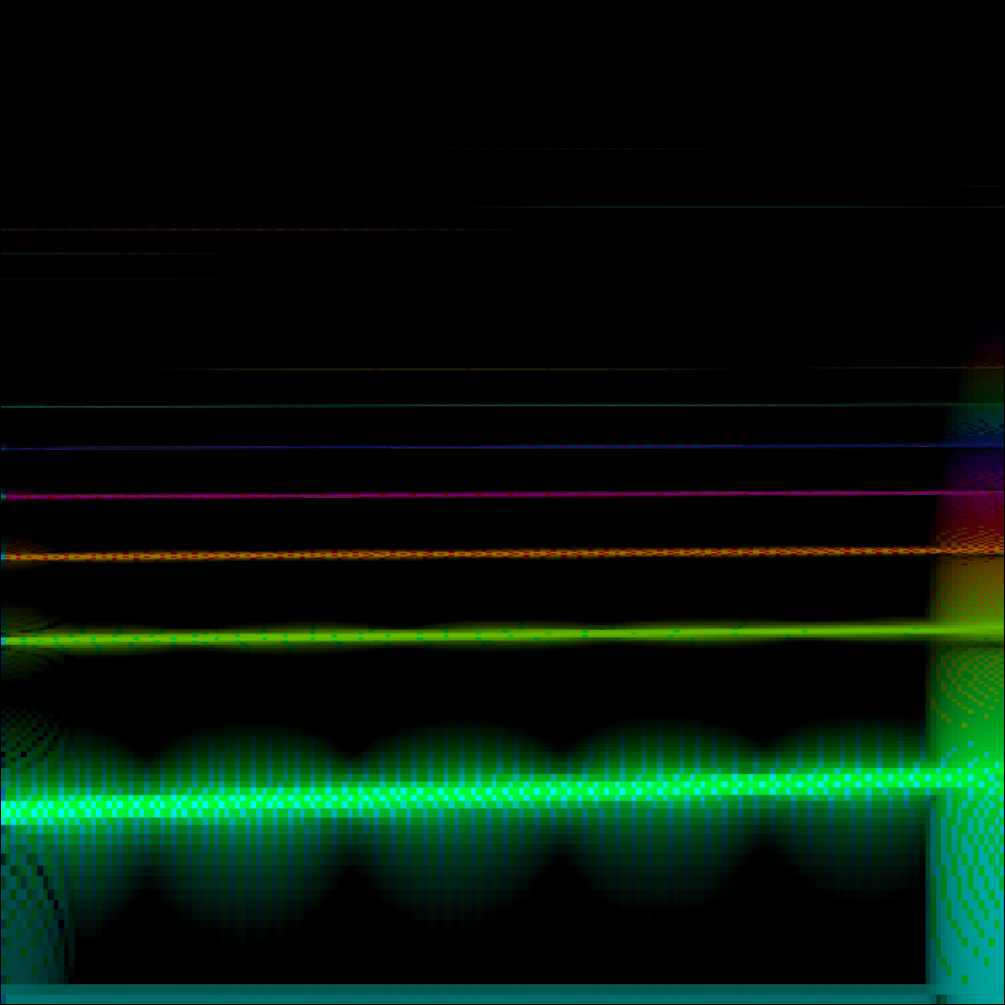}}
        \centerline{\footnotesize (a) Modal}
    \end{minipage}
	\begin{minipage}[b]{0.16\linewidth}
        \centering
        \centerline{\includegraphics[width=\linewidth]{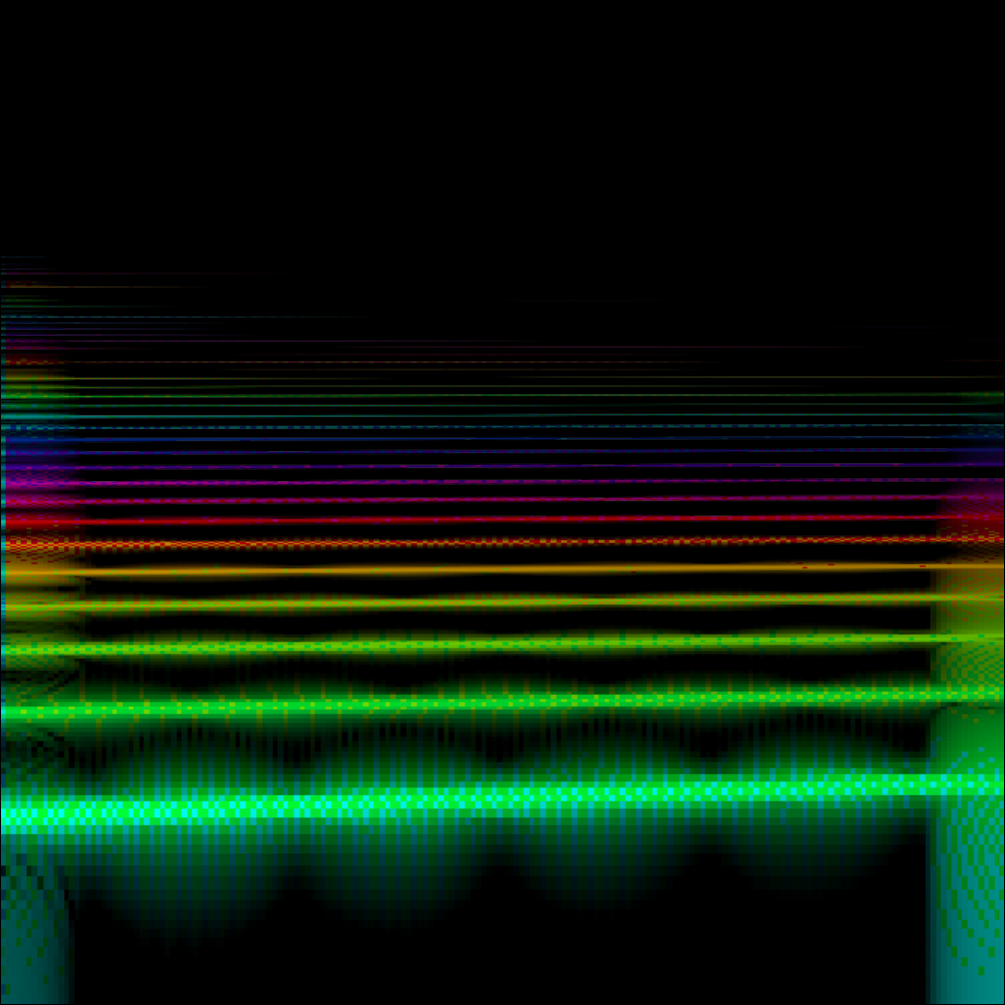}}
        \centerline{\footnotesize (b) DDSPish-{\scriptsize XFM}}
    \end{minipage}
	\begin{minipage}[b]{0.16\linewidth}
        \centering
        \centerline{\includegraphics[width=\linewidth]{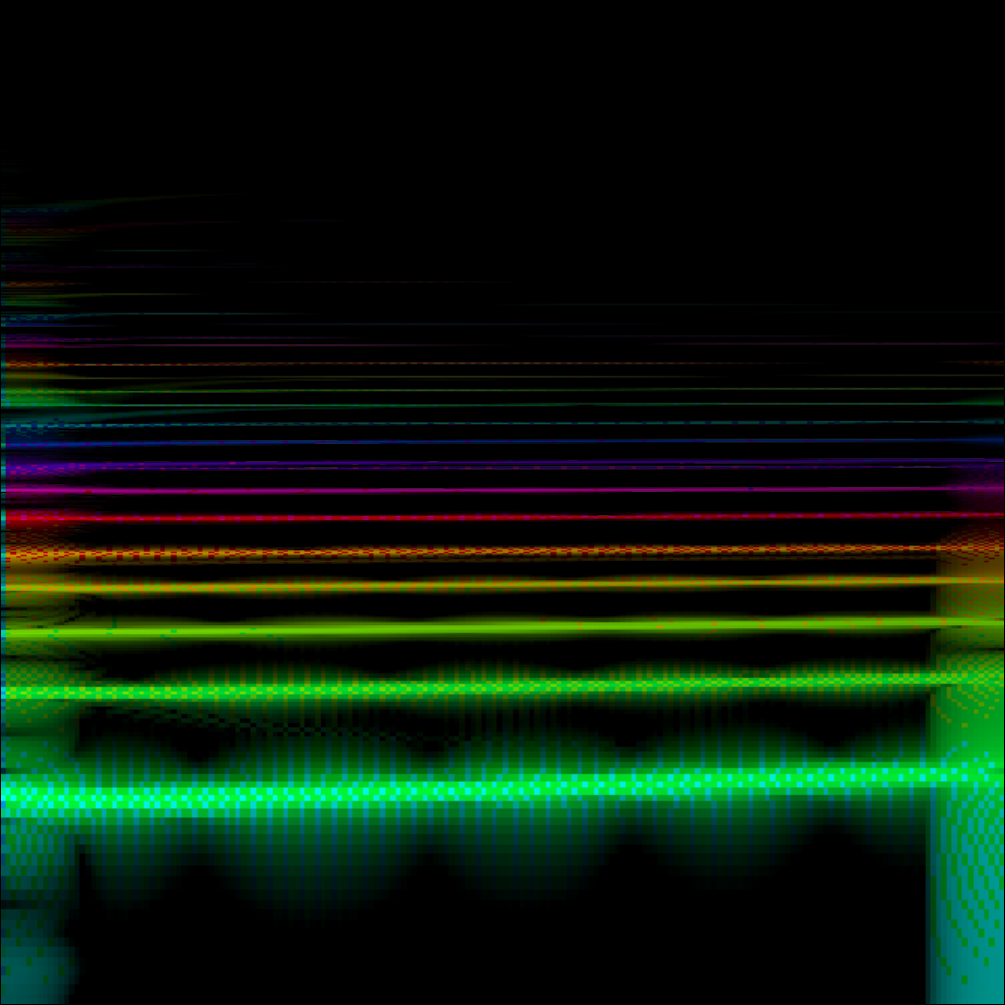}}
        \centerline{\footnotesize (c) DDSPish}
    \end{minipage}
	\begin{minipage}[b]{0.16\linewidth}
        \centering
        \centerline{\includegraphics[width=\linewidth]{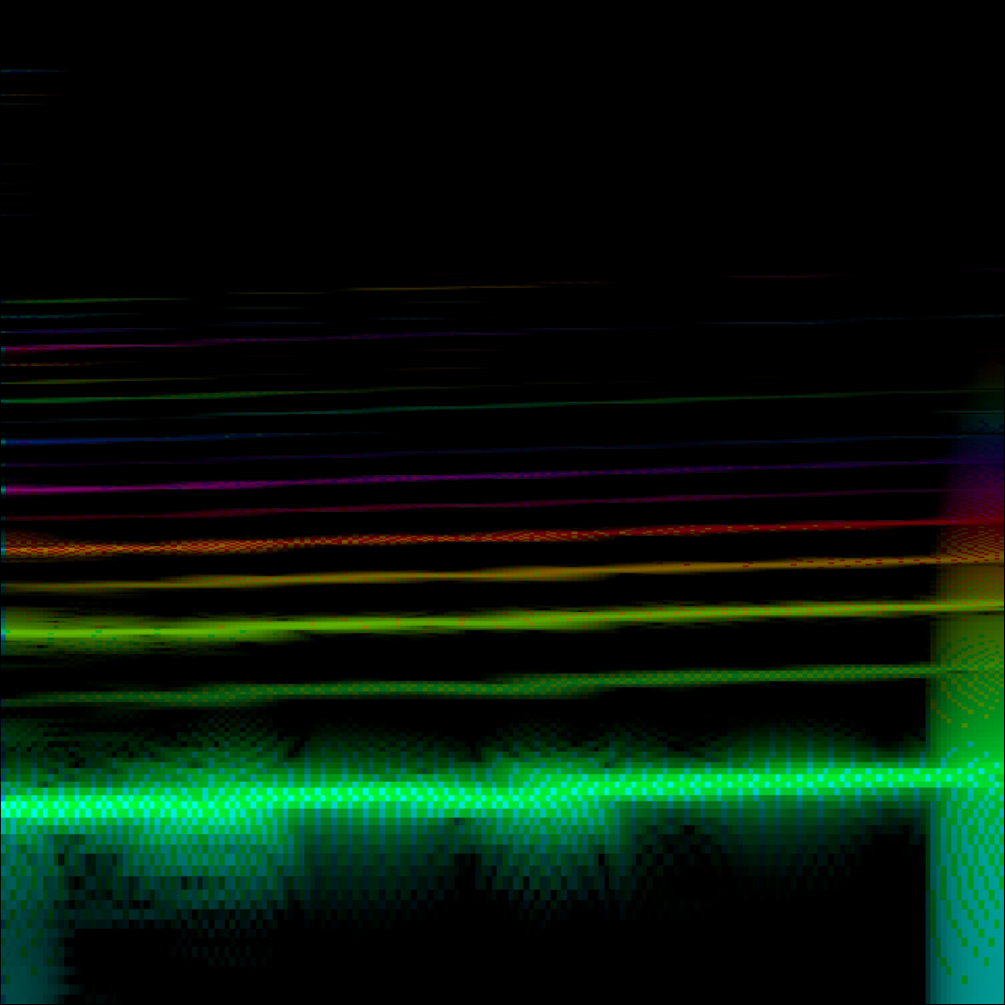}}
        \centerline{\footnotesize (d) DMSP}
    \end{minipage}
	\begin{minipage}[b]{0.16\linewidth}
        \centering
        \centerline{\includegraphics[width=\linewidth]{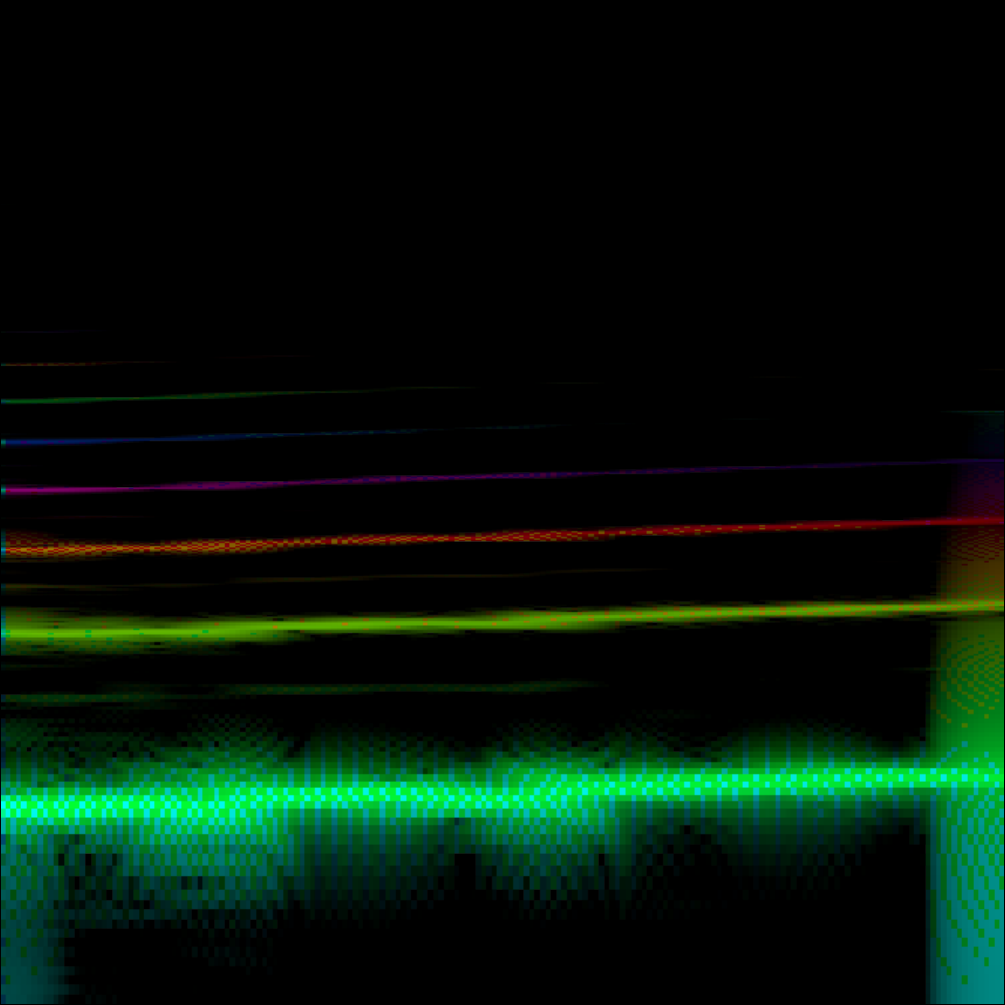}}
        \centerline{\footnotesize (e) DMSP-Hybrid}
    \end{minipage}
	\begin{minipage}[b]{0.16\linewidth}
        \centering
        \centerline{\includegraphics[width=\linewidth]{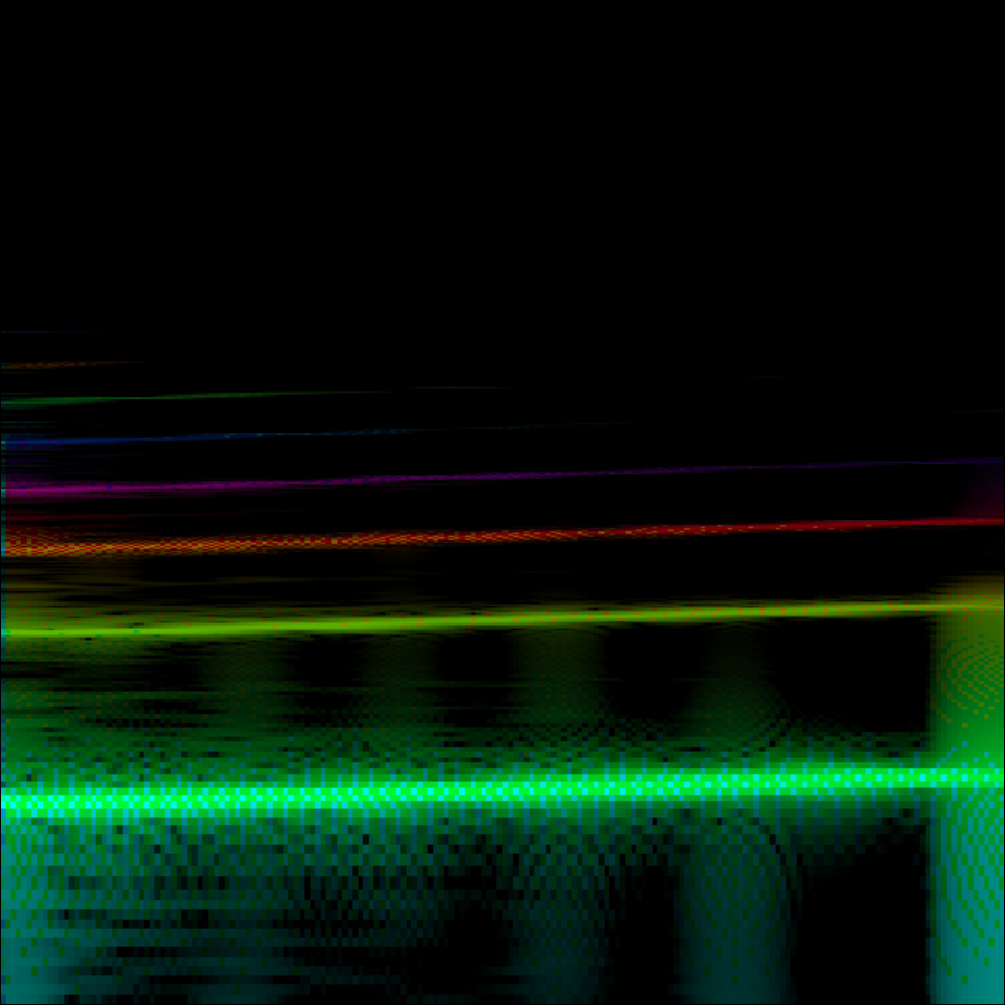}}
        \centerline{\footnotesize (f) FDTD (GT)}
    \end{minipage}
    \caption{Spectrogram of the synthesized samples on the test set.}
	\label{fig: rainbowgram}
\end{figure}

\begin{figure}
	\begin{minipage}[b]{0.16\linewidth}
        \centering
        \centerline{\includegraphics[width=\linewidth]{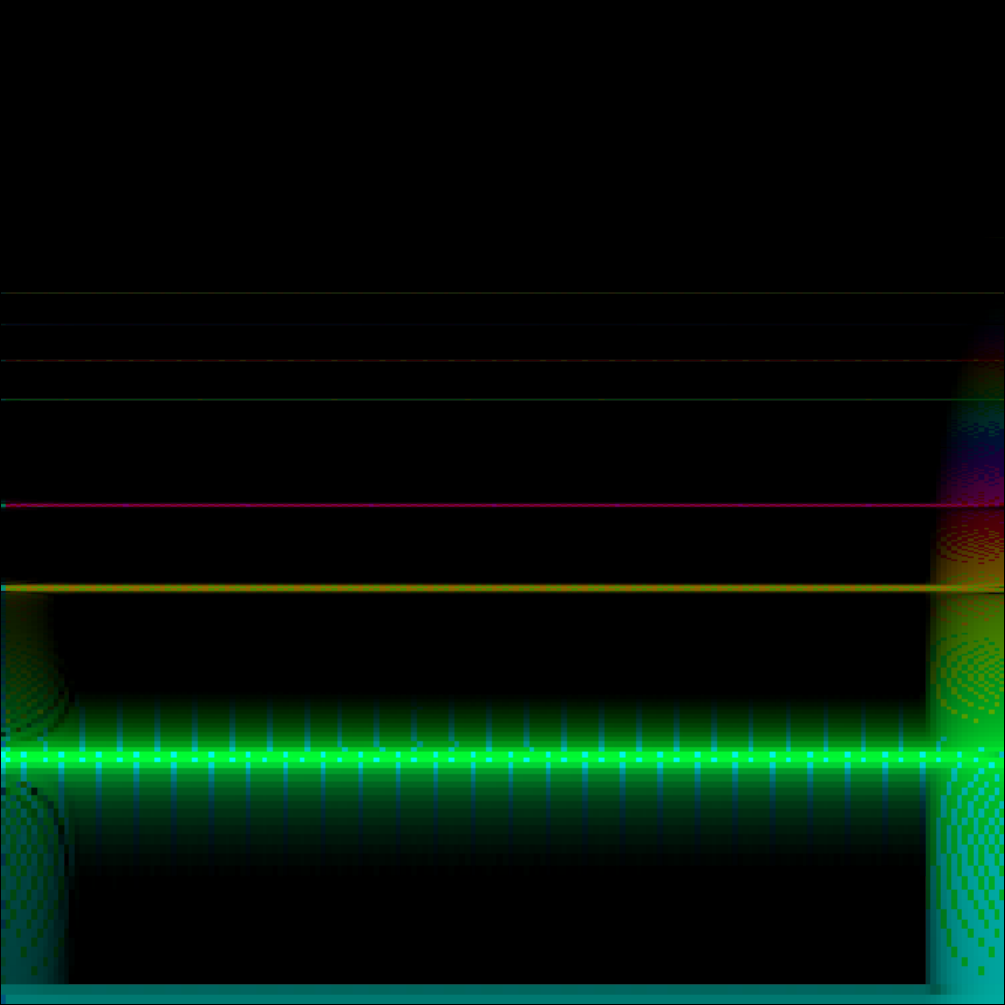}}
    \end{minipage}
	\begin{minipage}[b]{0.16\linewidth}
        \centering
        \centerline{\includegraphics[width=\linewidth]{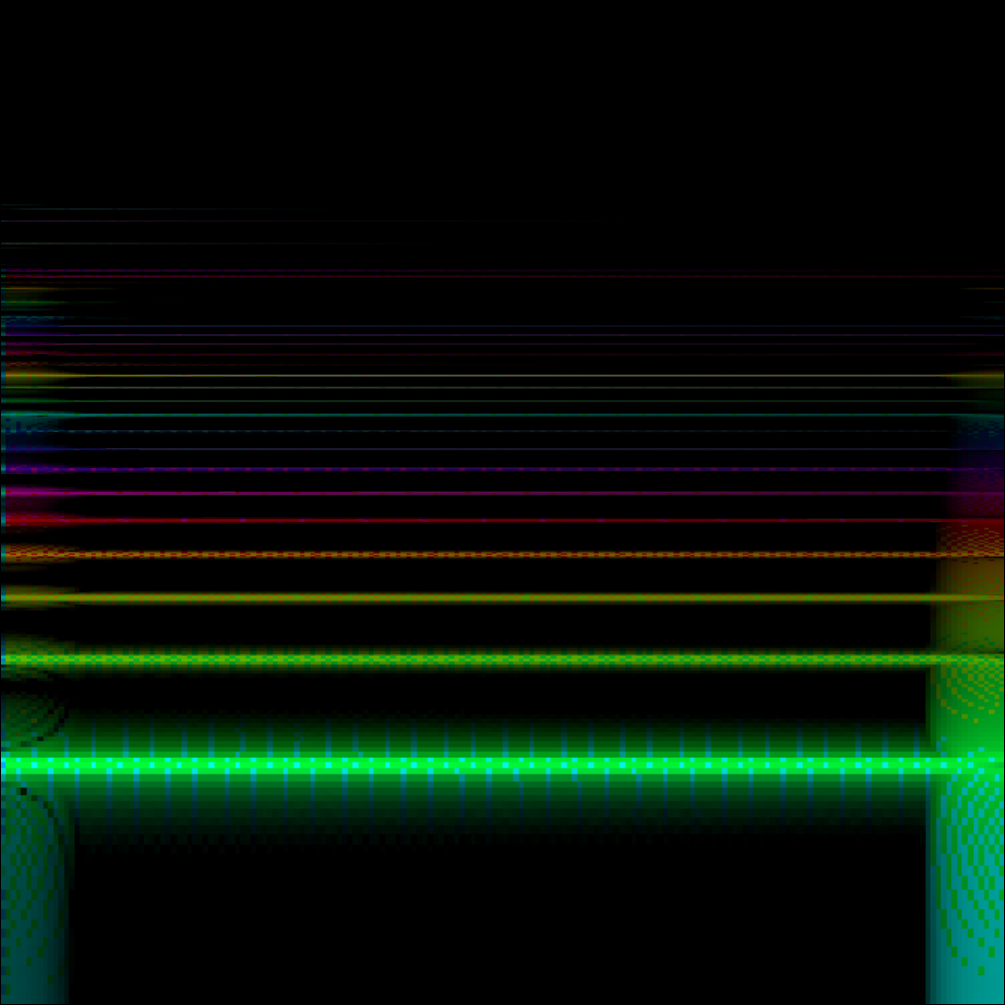}}
    \end{minipage}
	\begin{minipage}[b]{0.16\linewidth}
        \centering
        \centerline{\includegraphics[width=\linewidth]{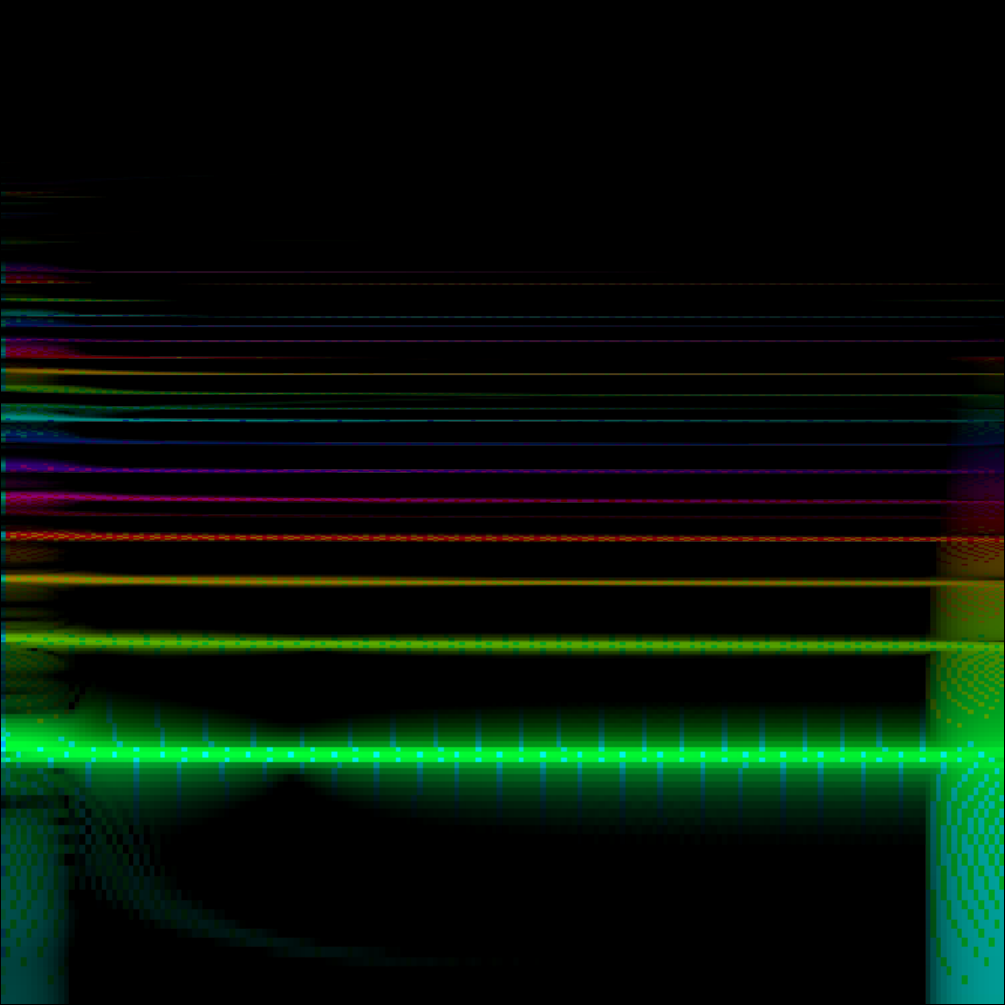}}
    \end{minipage}
	\begin{minipage}[b]{0.16\linewidth}
        \centering
        \centerline{\includegraphics[width=\linewidth]{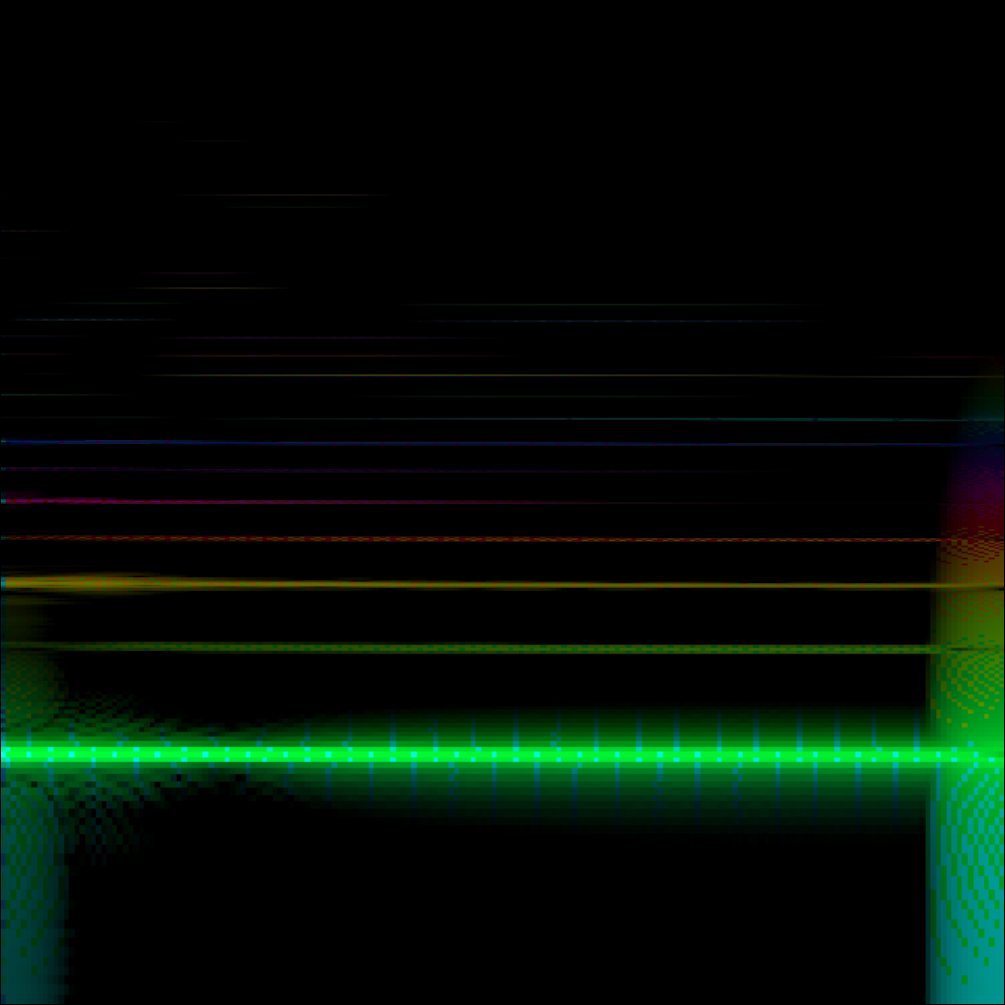}}
    \end{minipage}
	\begin{minipage}[b]{0.16\linewidth}
        \centering
        \centerline{\includegraphics[width=\linewidth]{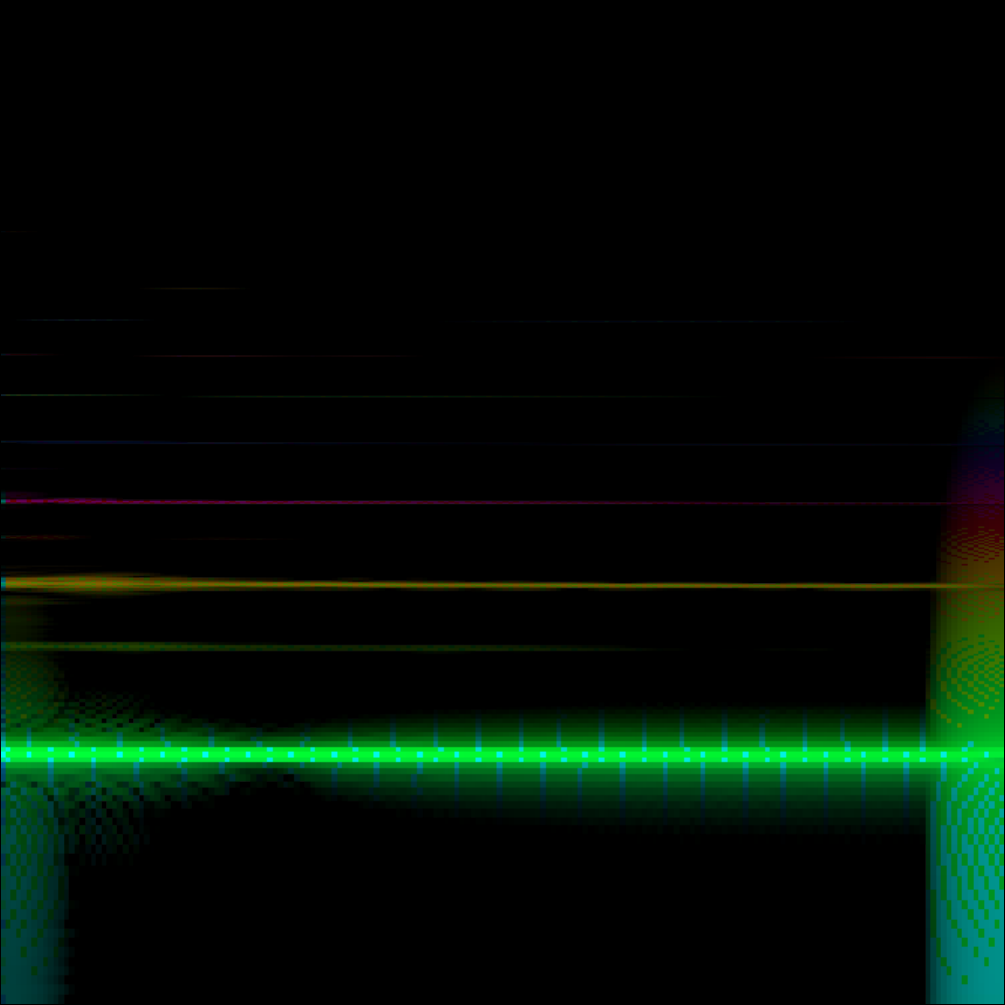}}
    \end{minipage}
	\begin{minipage}[b]{0.16\linewidth}
        \centering
        \centerline{\includegraphics[width=\linewidth]{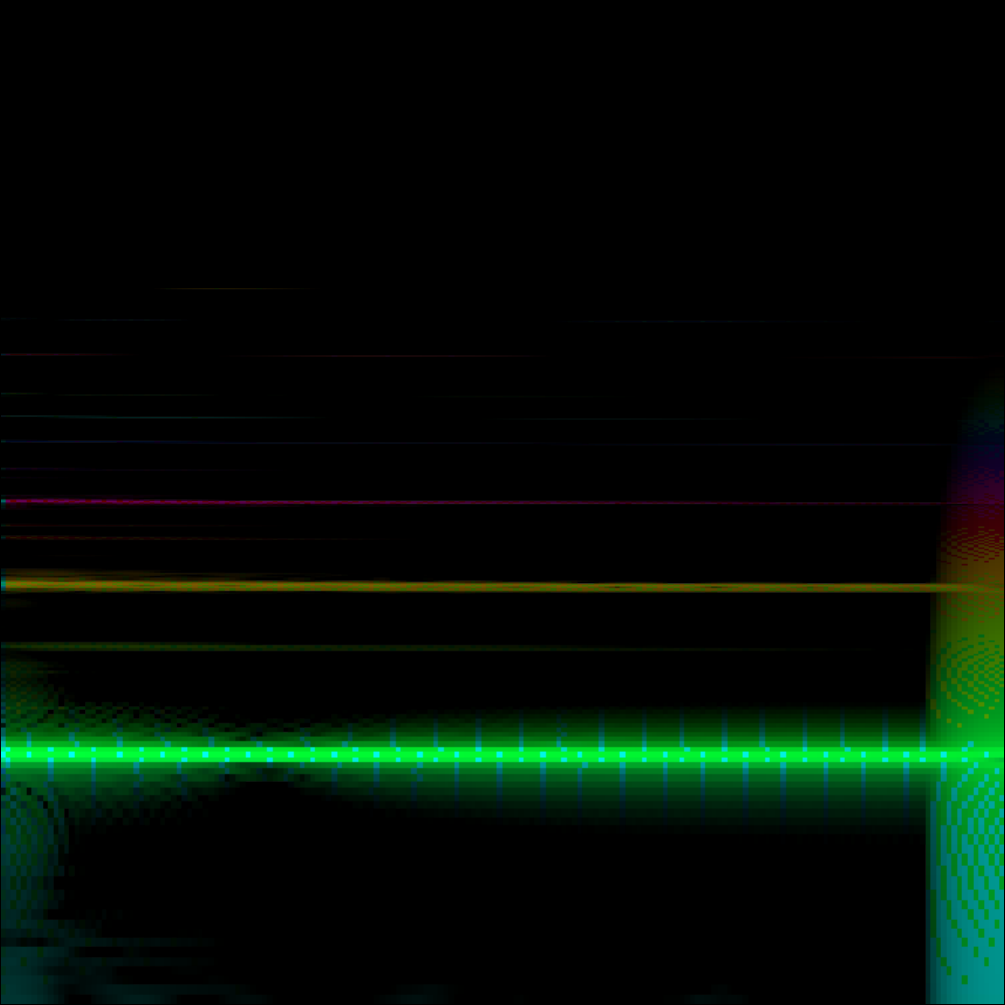}}
    \end{minipage}
	\begin{minipage}[b]{0.162\linewidth}
        \centering
        \centerline{\includegraphics[width=\linewidth]{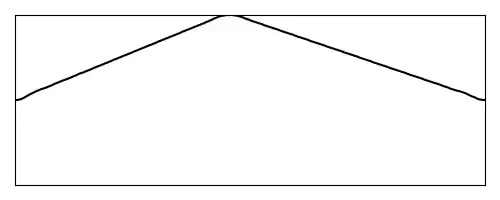}}
        \centerline{\footnotesize (a) Modal}
    \end{minipage}
	\begin{minipage}[b]{0.162\linewidth}
        \centering
        \centerline{\includegraphics[width=\linewidth]{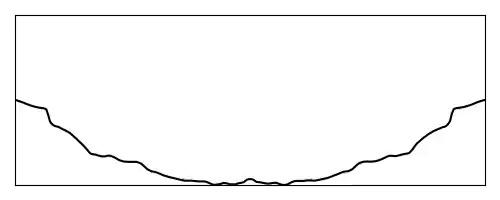}}
        \centerline{\footnotesize (b) DDSPish-{\scriptsize XFM}}
    \end{minipage}
	\begin{minipage}[b]{0.162\linewidth}
        \centering
        \centerline{\includegraphics[width=\linewidth]{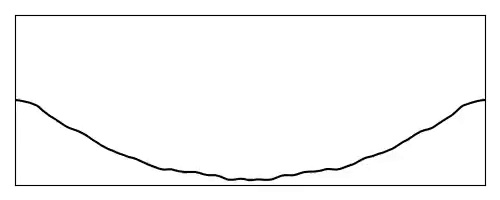}}
        \centerline{\footnotesize (c) DDSPish}
    \end{minipage}
	\begin{minipage}[b]{0.162\linewidth}
        \centering
        \centerline{\includegraphics[width=\linewidth]{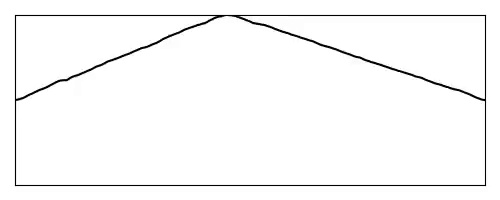}}
        \centerline{\footnotesize (d) DMSP}
    \end{minipage}
	\begin{minipage}[b]{0.162\linewidth}
        \centering
        \centerline{\includegraphics[width=\linewidth]{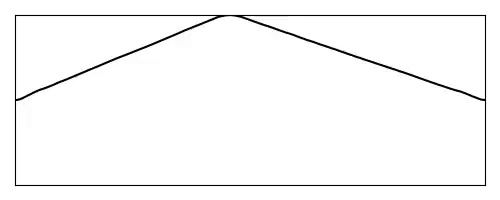}}
        \centerline{\footnotesize (e) DMSP-Hybrid}
    \end{minipage}
	\begin{minipage}[b]{0.16\linewidth}
        \centering
        \centerline{\includegraphics[width=\linewidth]{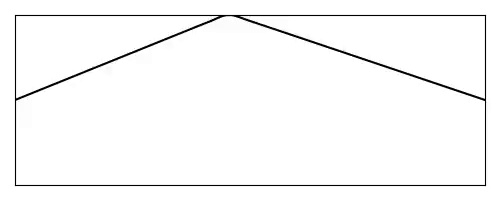}}
        \centerline{\footnotesize (f) FDTD (GT)}
    \end{minipage}
    \caption{Spectrograms and state samples of the synthesized samples on the test set. For the spectrograms shown in the first column, the intensity of the frequency (vertical axis) component for time (horizontal axis) is expressed as brightness, and for the states shown in the second column, the displacement (vertical axis) for space (horizontal axis).}
	\label{fig: initspec}
\end{figure}

\newpage
\textbf{String Sound Synthesis.}
Spectrograms of the test samples are visualized in \autoref{fig: rainbowgram} and \autoref{fig: initspec}.
For the spectrograms, the instantaneous frequencies are identified in a rainbow color map, where the color intensities represent the logarithmic magnitude of the power spectra.
Observing from \autoref{fig: rainbowgram}, the FDTD-simulated spectrogram clearly shows pitch glide and phantom partials at the beginning of the pluck.
In contrast, modal synthesis methods that model linear solutions do not show these nonlinear characteristics.
The DDSPish-{\scriptsize XFM} model employs a harmonic pitch skeleton comprising integer multiples of $f_0$, thereby precluding the inharmoicities resulting from stiffness.
The DDSPish model demonstrates enhanced mode estimation capabilities through learned FM, which modulates the harmonic pitch skeleton to be inharmonic.
However, there is scope for further improvement in frequency estimation, particularly in instances where the FM learning process is unstable for high frequencies.
On the other hand, the DMSP model, which estimates the mode frequency and amplitude from $u_0$, shows an improved pitch skeleton and stable frequency estimation.
The DMSP-Hybrid model, which learns to AM and FM the sinusoidal oscillators of the modal solution that requires mode precomputation, shows the most similar results to FDTD. 

\autoref{fig: initspec} shows a similar trend for the spectrograms.
In this example, however, we reveal a key difference that is not apparent in the spectrograms, but is very important for physical modeling:
It is the measure of displacement to space, namely the state plot.
The timestep for the state plot can be arbitrary, but we show the initial state for ease of comparison.
For methods that do not have a separate method for accurately predicting the mode information, unlike DMSP, difficulties can be found in accurately predicting the state.
This partially explains the SI-SDR scores for DDSPish and DDSPish-{\scriptsize XFM} in \autoref{tab: synthesis}.
Considering that the major difference between DMSP and DDSPish is whether the input frequency of the FM block is the mode frequency or an integer multiple, it can be inferred that accurately estimating the mode frequency is critical from a physical modeling perspective to fit the displacement of the strings over time.

\begin{table}
  \caption{Ablation Study}
  \label{tab: ablation}
  \centering
  \begin{tabular}{l cccc cccc}
    \toprule
    \multirow{3}[2]{*}{\textbf{Model}} & \multicolumn{4}{c}{\textbf{Linear}} & \multicolumn{4}{c}{\textbf{Nonlinear}} \\
    \cmidrule(r){2-5} \cmidrule(r){6-9}
    & \textbf{SI-SDR} & \textbf{SDR}  & \textbf{MSS} & \textbf{Pitch} 
    & \textbf{SI-SDR} & \textbf{SDR}  & \textbf{MSS} & \textbf{Pitch} \\
    & (dB, $\uparrow$)  & (dB, $\uparrow$)	 & (dB, $\downarrow$) & (Hz, $\downarrow$) 
    & (dB, $\uparrow$)  & (dB, $\uparrow$)	 & (dB, $\downarrow$) & (Hz, $\downarrow$) \\
    \midrule
    \textbf{DMSP-Hybrid}
       	& --2.844   & 1.496  & 12.525    & 0.792
        & 15.670    & 16.455 & 4.772     & 1.027  \\
    \quad w.o. $\mathcal{L}_{f_0}$
        & --2.919    & 0.774     & 13.487    & 0.792 
        & --5.418    & 1.509     & 8.983     & 2.653 \\
    \midrule
    \textbf{DMSP}
        & --22.298   & --2.000   & 12.504   & 1.717
        & --10.315   &   0.221   & 5.656    & 1.437 \\
    \quad w.o. $\mathcal{L}_{f_0}$
        & --21.351   & --2.699    & 13.482    & 1.717 
        & --16.435   & --1.074    & 9.060     & 2.922 \\
    \bottomrule
  \end{tabular}
\end{table}

\textbf{Ablation Study.}
The ablation study on the choice of the training loss functions is presented in \autoref{tab: ablation}.
Overall, for linear strings, the performance does not vary much depending on which loss is used, especially for pitch.
This is due to a gating applied to the FM block, which is designed to prevent FM from occurring when $\alpha$ is 1.
More specifically, we apply a gating, \textit{e.g.}, $\tanh(\alpha - 1)$, in such a way that frequency modulation only occurs when the value of $\alpha$ deviates from 1, which can directly affect pitch, and otherwise forces the mode estimation result to be used as is.
This is the main factor that determines the nonlinearity of the string.
For AM, there is no such masking depending on $\alpha$, so the remaining metrics except pitch do vary.
For nonlinear data, the training results vary depending on the loss design.
In particular, for all metrics, using $\mathcal{L}_{f_0}$ loss significantly improved performance over not using it.
This difference is especially noticeable in the MSS scores between DMSP-Hybrid w.o. $\mathcal{L}_{f_0}$ and DMSP,
where DMSP even gives better results over the model that uses precomputed mode frequencies for nonlinear MSS scores if $\mathcal{L}_{f_0}$ is applied.
This result reaffirms the validity of the FM block for nonlinear strings in terms of predicting dynamically varying frequencies such as pitch glide and shows that $\mathcal{L}_{f_0}$ loss is the loss function to train it effectively.

\begin{figure}
	\begin{minipage}[b]{0.49\linewidth}
        \centering
        \centerline{\includegraphics[width=\linewidth]{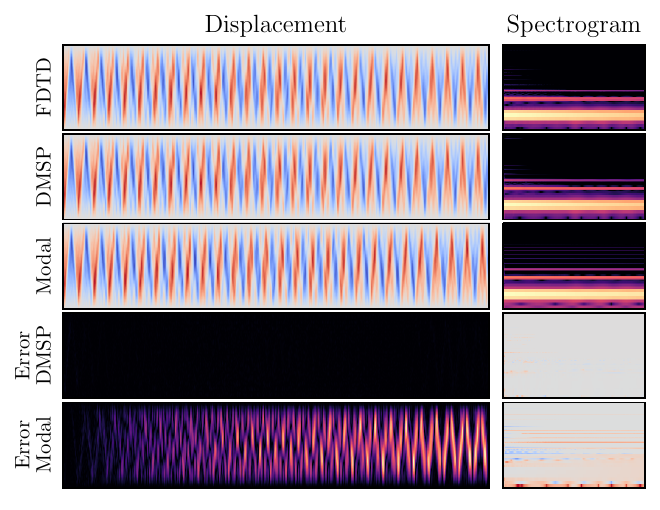}}
    \end{minipage}
	\begin{minipage}[b]{0.49\linewidth}
        \centering
        \centerline{\includegraphics[width=\linewidth]{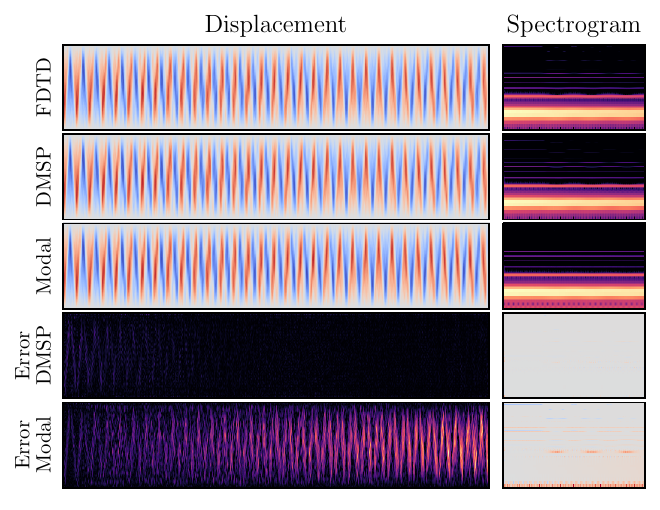}}
    \end{minipage}
    \vspace{-3mm}
    \caption{Simulated string state visualization.}
    \label{fig: state sample}
    \vspace{-5mm}
\end{figure}

\textbf{Motion Synthesis.}
The main advantage of DMSP, compared to the existing sound synthesis models, is that it can synthesize not only sound but also motion, which is one of the main characteristics of physical modeling techniques.
In particular, the DMSP can visualize the corresponding string motion as a video, although the receptive field and computational complexity required to obtain a solution \(u(x,t)\) for a single spatio-temporal point is of order 1, as shown in \autoref{tab: comparison}, when these solutions are pooled for a given \(x\in\Omega\) and \(t\in[0,1)\), the corresponding string motion can be visualized as a video.
\autoref{fig: state sample} visualizes the resulting transverse displacement of the string over time (horizontal axis) and space (vertical axis). The transverse displacement of the FDTD is coupled to the longitudinal motion, which is why it differs from the Modal synthesis output, which synthesizes the motion of a linear damped stiff string. The results output by DMSP show improved accuracy.

\section{Conclusion}
We present a novel neural network-based method that efficiently simulates plucked string motions. Our differentiable modal synthesis for physical modeling (DMSP) can simulate a dynamic nonlinear string motion by synthesizing the sound using the physical properties of the string. It is an efficient approximation of existing physical modeling methods.
We demonstrate the efficacy of training the neural network using mode frequency information by extending the DDSP with a modal synthesis pipeline. This opens the door to a new field of differentiable audio signal processing, extending it to the field of physical modeling for musical sound synthesis. While the proposed method offers control over several physical parameters of a musical instrument, it still faces limitations in terms of generalizing to physical parameters and sounds in real-world measurements.
This study paves the way for future research in this area. To the best of our knowledge, this is the first study to simultaneously synthesize sound and motion from the properties of a stringed instrument.

\newpage

\section*{Acknowledgments}
This work was partly supported by the National Research Foundation of Korea (NRF) grant funded by the Korea government (MSIT) [NO. RS-2023-00219429] and partly by Institute of Information \& communications Technology Planning \& Evaluation (IITP) grant funded by the Korea government (MSIT) [NO.RS-2021-II211343, Artificial Intelligence Graduate School Program (Seoul National University)].

\bibliographystyle{unsrtnat}
\bibliography{references}

\begin{thebibliography}{66}
\providecommand{\natexlab}[1]{#1}
\providecommand{\url}[1]{\texttt{#1}}
\expandafter\ifx\csname urlstyle\endcsname\relax
  \providecommand{\doi}[1]{doi: #1}\else
  \providecommand{\doi}{doi: \begingroup \urlstyle{rm}\Url}\fi

\bibitem[Rayleigh(1896)]{rayleigh1896theory}
John William Strutt~Baron Rayleigh.
\newblock \emph{The theory of sound}, volume~2.
\newblock Macmillan, 1896.

\bibitem[Donkin(1884)]{donkin1884acoustics}
William~Fishburn Donkin.
\newblock \emph{Acoustics: Theoretical. Part 1}.
\newblock Clarendon Press, 1884.

\bibitem[Fletcher(1964)]{fletcher1964normal}
Harvey Fletcher.
\newblock Normal vibration frequencies of a stiff piano string.
\newblock \emph{The Journal of the Acoustical Society of America}, 36\penalty0 (1):\penalty0 203--209, 1964.

\bibitem[Ruiz(1970)]{ruiz1970technique}
Pierre~Michel Ruiz.
\newblock \emph{A technique for simulating the vibration of strings with a digital computer}.
\newblock PhD thesis, University of Illinois at Urbana-Champaign, 1970.

\bibitem[Hiller and Ruiz(1971)]{hiller1971synthesizing}
Lejaren Hiller and Pierre Ruiz.
\newblock Synthesizing musical sounds by solving the wave equation for vibrating objects: Part 1.
\newblock \emph{Journal of the Audio Engineering Society}, 19\penalty0 (6), 1971.

\bibitem[Schwarz(2007)]{schwarz2007corpus}
Diemo Schwarz.
\newblock Corpus-based concatenative synthesis.
\newblock \emph{IEEE signal processing magazine}, 24\penalty0 (2):\penalty0 92--104, 2007.

\bibitem[Serra and Smith(1990)]{serra1990spectral}
Xavier Serra and Julius Smith.
\newblock Spectral modeling synthesis: A sound analysis/synthesis system based on a deterministic plus stochastic decomposition.
\newblock \emph{Computer Music Journal}, 14\penalty0 (4):\penalty0 12--24, 1990.

\bibitem[Smith~III(2010)]{smith2010physical}
Julius~O Smith~III.
\newblock \emph{Physical audio signal processing: For virtual musical instruments and audio effects}.
\newblock 2010.

\bibitem[Fettweis(1986)]{fettweis1986wave}
Alfred Fettweis.
\newblock Wave digital filters: Theory and practice.
\newblock \emph{Proceedings of the IEEE}, 74\penalty0 (2):\penalty0 270--327, 1986.

\bibitem[Bilbao(2009)]{bilbao2009numerical}
Stefan Bilbao.
\newblock \emph{Numerical sound synthesis: finite difference schemes and simulation in musical acoustics}.
\newblock John Wiley \& Sons, 2009.

\bibitem[Hayes et~al.(2024)Hayes, Shier, Fazekas, McPherson, and Saitis]{hayes2024review}
Ben Hayes, Jordie Shier, Gy{\"o}rgy Fazekas, Andrew McPherson, and Charalampos Saitis.
\newblock A review of differentiable digital signal processing for music and speech synthesis.
\newblock \emph{Frontiers in Signal Processing}, 3:\penalty0 1284100, 2024.

\bibitem[van~den Oord et~al.(2016)van~den Oord, Dieleman, Zen, Simonyan, Vinyals, Graves, Kalchbrenner, Senior, and Kavukcuoglu]{wavenet}
Aäron van~den Oord, Sander Dieleman, Heiga Zen, Karen Simonyan, Oriol Vinyals, Alexander Graves, Nal Kalchbrenner, Andrew Senior, and Koray Kavukcuoglu.
\newblock Wavenet: A generative model for raw audio.
\newblock In \emph{Arxiv}, 2016.
\newblock URL \url{https://arxiv.org/abs/1609.03499}.

\bibitem[Donahue et~al.(2018)Donahue, McAuley, and Puckette]{donahue2018adversarial}
Chris Donahue, Julian McAuley, and Miller Puckette.
\newblock Adversarial audio synthesis.
\newblock In \emph{International Conference on Learning Representations}, 2018.

\bibitem[Engel et~al.(2018)Engel, Agrawal, Chen, Gulrajani, Donahue, and Roberts]{engel2018gansynth}
Jesse Engel, Kumar~Krishna Agrawal, Shuo Chen, Ishaan Gulrajani, Chris Donahue, and Adam Roberts.
\newblock Gansynth: Adversarial neural audio synthesis.
\newblock In \emph{International Conference on Learning Representations}, 2018.

\bibitem[Valin and Skoglund(2019)]{valin2019lpcnet}
Jean-Marc Valin and Jan Skoglund.
\newblock Lpcnet: Improving neural speech synthesis through linear prediction.
\newblock In \emph{ICASSP 2019-2019 IEEE International Conference on Acoustics, Speech and Signal Processing (ICASSP)}, pages 5891--5895. IEEE, 2019.

\bibitem[Engel et~al.(2019)Engel, Gu, Roberts, et~al.]{engel2019ddsp}
Jesse Engel, Chenjie Gu, Adam Roberts, et~al.
\newblock Ddsp: Differentiable digital signal processing.
\newblock In \emph{International Conference on Learning Representations}, 2019.

\bibitem[Wang et~al.(2019)Wang, Takaki, and Yamagishi]{wang2019neural}
Xin Wang, Shinji Takaki, and Junichi Yamagishi.
\newblock Neural source-filter waveform models for statistical parametric speech synthesis.
\newblock \emph{IEEE/ACM Transactions on Audio, Speech, and Language Processing}, 28:\penalty0 402--415, 2019.

\bibitem[Hayes et~al.(2021)Hayes, Saitis, and Fazekas]{hayes2021neural}
Ben Hayes, Charalampos Saitis, and Gy{\"o}rgy Fazekas.
\newblock Neural waveshaping synthesis.
\newblock \emph{arXiv preprint arXiv:2107.05050}, 2021.

\bibitem[Caspe et~al.(2022)Caspe, McPherson, and Sandler]{caspe2022ddx7}
Franco Caspe, Andrew McPherson, and Mark Sandler.
\newblock {DDX7: Differentiable FM Synthesis of Musical Instrument Sounds}.
\newblock \emph{Proceedings of the 23rd International Society for Music Information Retrieval Conference}, 2022.

\bibitem[Braun(2023)]{Braun_DX7-JAX_2023}
David Braun.
\newblock {DX7-JAX}, November 2023.
\newblock URL \url{https://github.com/DBraun/DX7-JAX}.

\bibitem[Wu et~al.(2021)Wu, Manilow, Deng, Swavely, Kastner, Cooijmans, Courville, Huang, and Engel]{wu2021midi}
Yusong Wu, Ethan Manilow, Yi~Deng, Rigel Swavely, Kyle Kastner, Tim Cooijmans, Aaron Courville, Cheng-Zhi~Anna Huang, and Jesse Engel.
\newblock Midi-ddsp: Detailed control of musical performance via hierarchical modeling.
\newblock In \emph{International Conference on Learning Representations}, 2021.

\bibitem[Choi et~al.(2021)Choi, Lee, Kim, Lee, Heo, and Lee]{choi2021neural}
Hyeong-Seok Choi, Juheon Lee, Wansoo Kim, Jie Lee, Hoon Heo, and Kyogu Lee.
\newblock Neural analysis and synthesis: Reconstructing speech from self-supervised representations.
\newblock \emph{Advances in Neural Information Processing Systems}, 34:\penalty0 16251--16265, 2021.

\bibitem[Choi et~al.(2022)Choi, Yang, Lee, and Kim]{choi2022nansy}
Hyeong-Seok Choi, Jinhyeok Yang, Juheon Lee, and Hyeongju Kim.
\newblock Nansy++: Unified voice synthesis with neural analysis and synthesis.
\newblock In \emph{The Eleventh International Conference on Learning Representations}, 2022.

\bibitem[Barahona-R{\'\i}os and Collins(2024)]{barahona2024noisebandnet}
Adri{\'a}n Barahona-R{\'\i}os and Tom Collins.
\newblock Noisebandnet: controllable time-varying neural synthesis of sound effects using filterbanks.
\newblock \emph{IEEE/ACM Transactions on Audio, Speech, and Language Processing}, 32:\penalty0 1573--1585, 2024.

\bibitem[Renault et~al.(2022)Renault, Mignot, and Roebel]{renault2022differentiable}
Lenny Renault, R{\'e}mi Mignot, and Axel Roebel.
\newblock Differentiable piano model for midi-to-audio performance synthesis.
\newblock In \emph{25th International Conference on Digital Audio Effects (DAFx20in22)}, 2022.

\bibitem[Rigaud et~al.(2011)Rigaud, David, and Daudet]{rigaud2011parametric}
Fran{\c{c}}ois Rigaud, Bertrand David, and Laurent Daudet.
\newblock A parametric model of piano tuning.
\newblock In \emph{Proc. of the 14th Int. Conf. on Digital Audio Effects (DAFx-11)}, pages 393--399, 2011.

\bibitem[Henningsson and Team(2011)]{henningsson2011fluidsynth}
David Henningsson and FD~Team.
\newblock Fluidsynth real-time and thread safety challenges.
\newblock In \emph{Proceedings of the 9th International Linux Audio Conference, Maynooth University, Ireland}, pages 123--128, 2011.

\bibitem[Bank and Chabassier(2018)]{bank2018model}
Balazs Bank and Juliette Chabassier.
\newblock Model-based digital pianos: from physics to sound synthesis.
\newblock \emph{IEEE Signal Processing Magazine}, 36\penalty0 (1):\penalty0 103--114, 2018.

\bibitem[Schlecht et~al.(2022)Schlecht, Parker, Sch{\"a}fer, and Rabenstein]{schlecht2022physical}
Sebastian Schlecht, Julian Parker, Maximilian Sch{\"a}fer, and Rudolf Rabenstein.
\newblock Physical modeling using recurrent neural networks with fast convolutional layers.
\newblock In \emph{International Conference on Digital Audio Effects}, pages 138--145. DAFx, 2022.

\bibitem[James et~al.(2006)James, Barbi{\v{c}}, and Pai]{james2006precomputed}
Doug~L James, Jernej Barbi{\v{c}}, and Dinesh~K Pai.
\newblock Precomputed acoustic transfer: output-sensitive, accurate sound generation for geometrically complex vibration sources.
\newblock \emph{ACM Transactions on Graphics (TOG)}, 25\penalty0 (3):\penalty0 987--995, 2006.

\bibitem[Wang and James(2019)]{wang2019kleinpat}
Jui-Hsien Wang and Doug~L James.
\newblock Kleinpat: optimal mode conflation for time-domain precomputation of acoustic transfer.
\newblock \emph{ACM Trans. Graph.}, 38\penalty0 (4):\penalty0 122--1, 2019.

\bibitem[O'Brien et~al.(2002)O'Brien, Shen, and Gatchalian]{o2002synthesizing}
James~F O'Brien, Chen Shen, and Christine~M Gatchalian.
\newblock Synthesizing sounds from rigid-body simulations.
\newblock In \emph{Proceedings of the 2002 ACM SIGGRAPH/Eurographics symposium on Computer animation}, pages 175--181, 2002.

\bibitem[Jin et~al.(2020)Jin, Li, Qu, Manocha, and Wang]{jin2020deep}
Xutong Jin, Sheng Li, Tianshu Qu, Dinesh Manocha, and Guoping Wang.
\newblock Deep-modal: real-time impact sound synthesis for arbitrary shapes.
\newblock In \emph{Proceedings of the 28th ACM International Conference on Multimedia}, pages 1171--1179, 2020.

\bibitem[Jin et~al.(2022)Jin, Li, Wang, and Manocha]{jin2022neuralsound}
Xutong Jin, Sheng Li, Guoping Wang, and Dinesh Manocha.
\newblock Neuralsound: learning-based modal sound synthesis with acoustic transfer.
\newblock \emph{ACM Transactions on Graphics (TOG)}, 41\penalty0 (4):\penalty0 1--15, 2022.

\bibitem[Diaz et~al.(2023)Diaz, Hayes, Saitis, Fazekas, and Sandler]{diaz2023rigid}
Rodrigo Diaz, Ben Hayes, Charalampos Saitis, Gy{\"o}rgy Fazekas, and Mark Sandler.
\newblock Rigid-body sound synthesis with differentiable modal resonators.
\newblock In \emph{ICASSP 2023-2023 IEEE International Conference on Acoustics, Speech and Signal Processing (ICASSP)}, pages 1--5. IEEE, 2023.

\bibitem[Kirchhoff(1897)]{kirchhoff1897vorlesungen}
Gustav Kirchhoff.
\newblock \emph{Vorlesungen {\"u}ber mechanik}, volume~1.
\newblock BG Teubner, 1897.

\bibitem[Carrier(1945)]{carrier1945non}
GF~Carrier.
\newblock On the non-linear vibration problem of the elastic string.
\newblock \emph{Quarterly of applied mathematics}, 3\penalty0 (2), 1945.

\bibitem[Bilbao and Ducceschi(2023)]{bilbao2023models}
Stefan Bilbao and Michele Ducceschi.
\newblock Models of musical string vibration.
\newblock \emph{Acoustical Science and Technology}, 2023.

\bibitem[Morse and Ingard(1986)]{morse1986theoretical}
Philip~McCord Morse and K~Uno Ingard.
\newblock \emph{Theoretical acoustics}.
\newblock Princeton university press, 1986.

\bibitem[Anand(1969)]{anand1969large}
GV~Anand.
\newblock Large-amplitude damped free vibration of a stretched string.
\newblock \emph{The Journal of the Acoustical Society of America}, 45\penalty0 (5):\penalty0 1089--1096, 1969.

\bibitem[Bilbao(2005)]{bilbao2005conservative}
Stefan Bilbao.
\newblock Conservative numerical methods for nonlinear strings.
\newblock \emph{The Journal of the Acoustical Society of America}, 118\penalty0 (5):\penalty0 3316--3327, 2005.

\bibitem[Bilbao(2004)]{bilbao2004energy}
Stefan Bilbao.
\newblock Energy-conserving finite difference schemes for tension-modulated strings.
\newblock In \emph{2004 IEEE International Conference on Acoustics, Speech, and Signal Processing}, volume~4, pages iv--iv. IEEE, 2004.

\bibitem[Maestre et~al.(2017)Maestre, Scavone, and Smith]{maestre2017joint}
Esteban Maestre, Gary~P Scavone, and Julius~O Smith.
\newblock Joint modeling of bridge admittance and body radiativity for efficient synthesis of string instrument sound by digital waveguides.
\newblock \emph{IEEE/ACM Transactions on Audio, Speech, and Language Processing}, 25\penalty0 (5):\penalty0 1128--1139, 2017.

\bibitem[Trautmann and Rabenstein(2004)]{trautmann2004multirate}
Lutz Trautmann and Rudolf Rabenstein.
\newblock Multirate simulations of string vibrations including nonlinear fret-string interactions using the functional transformation method.
\newblock \emph{EURASIP Journal on Advances in Signal Processing}, 2004:\penalty0 1--15, 2004.

\bibitem[Moin and Mahesh(1998)]{moin1998direct}
Parviz Moin and Krishnan Mahesh.
\newblock Direct numerical simulation: a tool in turbulence research.
\newblock \emph{Annual review of fluid mechanics}, 30\penalty0 (1):\penalty0 539--578, 1998.

\bibitem[Taflove et~al.(2013)Taflove, Oskooi, and Johnson]{taflove2013advances}
Allen Taflove, Ardavan Oskooi, and Steven~G Johnson.
\newblock \emph{Advances in FDTD computational electrodynamics: photonics and nanotechnology}.
\newblock Artech house, 2013.

\bibitem[Lee et~al.(2024)Lee, Choi, and Lee]{lee2024string}
Jin~Woo Lee, Min~Jun Choi, and Kyogu Lee.
\newblock String sound synthesizer on gpu-accelerated finite difference scheme.
\newblock In \emph{ICASSP 2024-2024 IEEE International Conference on Acoustics, Speech and Signal Processing (ICASSP)}, pages 1491--1495. IEEE, 2024.

\bibitem[Adrien(1991)]{adrien1991missing}
Jean-Marie Adrien.
\newblock The missing link: Modal synthesis.
\newblock In \emph{Representations of musical signals}, pages 269--298. 1991.

\bibitem[Adrien and Rodet(1985)]{adrien1985physical}
Jean-Marie Adrien and Xavier Rodet.
\newblock Physical models of instruments: A modular approach, application to strings.
\newblock In \emph{ICMC}, 1985.

\bibitem[Morrison and Adrien(1993)]{morrison1993mosaic}
Joseph~Derek Morrison and Jean-Marie Adrien.
\newblock Mosaic: A framework for modal synthesis.
\newblock \emph{Computer Music Journal}, 17\penalty0 (1):\penalty0 45--56, 1993.

\bibitem[Lee et~al.(2022)Lee, Choi, and Lee]{lee2022differentiable}
Sungho Lee, Hyeong-Seok Choi, and Kyogu Lee.
\newblock Differentiable artificial reverberation.
\newblock \emph{IEEE/ACM Transactions on Audio, Speech, and Language Processing}, 30:\penalty0 2541--2556, 2022.

\bibitem[Rahimi and Recht(2007)]{rahimi2007random}
Ali Rahimi and Benjamin Recht.
\newblock Random features for large-scale kernel machines.
\newblock \emph{Advances in neural information processing systems}, 20, 2007.

\bibitem[Tancik et~al.(2020)Tancik, Srinivasan, Mildenhall, Fridovich-Keil, Raghavan, Singhal, Ramamoorthi, Barron, and Ng]{tancik2020fourier}
Matthew Tancik, Pratul Srinivasan, Ben Mildenhall, Sara Fridovich-Keil, Nithin Raghavan, Utkarsh Singhal, Ravi Ramamoorthi, Jonathan Barron, and Ren Ng.
\newblock Fourier features let networks learn high frequency functions in low dimensional domains.
\newblock \emph{Advances in neural information processing systems}, 33:\penalty0 7537--7547, 2020.

\bibitem[Turian and Henry(2020)]{turian2020m}
Joseph Turian and Max Henry.
\newblock I’m sorry for your loss: Spectrally-based audio distances are bad at pitch.
\newblock In \emph{''I Can't Believe It's Not Better!''NeurIPS 2020 workshop}, 2020.

\bibitem[Hayes et~al.(2023)Hayes, Saitis, and Fazekas]{hayes2023sinusoidal}
Ben Hayes, Charalampos Saitis, and Gy{\"o}rgy Fazekas.
\newblock Sinusoidal frequency estimation by gradient descent.
\newblock In \emph{ICASSP 2023-2023 IEEE International Conference on Acoustics, Speech and Signal Processing (ICASSP)}, pages 1--5. IEEE, 2023.

\bibitem[Torres et~al.(2024)Torres, Peeters, and Richard]{torres2024unsupervised}
Bernardo Torres, Geoffroy Peeters, and Ga{\"e}l Richard.
\newblock Unsupervised harmonic parameter estimation using differentiable dsp and spectral optimal transport.
\newblock In \emph{ICASSP 2024-2024 IEEE International Conference on Acoustics, Speech and Signal Processing (ICASSP)}, pages 1176--1180. IEEE, 2024.

\bibitem[Schw{\"a}r and M{\"u}ller(2023)]{schwar2023multi}
Simon Schw{\"a}r and Meinard M{\"u}ller.
\newblock Multi-scale spectral loss revisited.
\newblock \emph{IEEE Signal Processing Letters}, 30:\penalty0 1712--1716, 2023.

\bibitem[Engel et~al.(2020)Engel, Swavely, Hantrakul, Roberts, and Hawthorne]{engel2020self}
Jesse Engel, Rigel Swavely, Lamtharn~Hanoi Hantrakul, Adam Roberts, and Curtis Hawthorne.
\newblock Self-supervised pitch detection by inverse audio synthesis.
\newblock In \emph{ICML 2020 Workshop on Self-supervision in Audio and Speech}, 2020.

\bibitem[Kim et~al.(2018)Kim, Salamon, Li, and Bello]{kim2018crepe}
Jong~Wook Kim, Justin Salamon, Peter Li, and Juan~Pablo Bello.
\newblock Crepe: A convolutional representation for pitch estimation.
\newblock In \emph{ICASSP}, pages 161--165. IEEE, 2018.

\bibitem[Le~Roux et~al.(2019)Le~Roux, Wisdom, Erdogan, and Hershey]{le2019sdr}
Jonathan Le~Roux, Scott Wisdom, Hakan Erdogan, and John~R Hershey.
\newblock Sdr--half-baked or well done?
\newblock In \emph{ICASSP 2019-2019 IEEE International Conference on Acoustics, Speech and Signal Processing (ICASSP)}, pages 626--630. IEEE, 2019.

\bibitem[Bank and Sujbert(2006)]{bank2006physics}
Bal{\'a}zs Bank and L{\'a}szl{\'o} Sujbert.
\newblock Physics-based sound synthesis of string instruments including geometric nonlinearities, 2006.

\bibitem[Bank and Sujbert(2004)]{bank2004piano}
Balaz Bank and Laszlo Sujbert.
\newblock A piano model including longitudinal string vibrations.
\newblock In \emph{Proceedings of the Digital Audio Effects Conference}, pages 89--94, 2004.

\bibitem[Trautmann and Rabenstein(2012)]{trautmann2012digital}
Lutz Trautmann and Rudolf Rabenstein.
\newblock \emph{Digital sound synthesis by physical modeling using the functional transformation method}.
\newblock Springer Science \& Business Media, 2012.

\bibitem[Sch{\"a}fer et~al.(2016)Sch{\"a}fer, Frenst{\'a}tsk{\`y}, and Rabenstein]{schafer2016physical}
Maximilian Sch{\"a}fer, Petr Frenst{\'a}tsk{\`y}, and Rudolf Rabenstein.
\newblock A physical string model with adjustable boundary conditions.
\newblock In \emph{19th International Conference on Digital Audio Effects (DAFx-16), Brno, Czech Republic}, pages 159--166, 2016.

\bibitem[Rabenstein et~al.(2009)Rabenstein, Koch, and Popp]{rabenstein2009tubular}
Rudolf Rabenstein, Tilman Koch, and Christian Popp.
\newblock Tubular bells: A physical and algorithmic model.
\newblock \emph{IEEE transactions on audio, speech, and language processing}, 18\penalty0 (4):\penalty0 881--890, 2009.

\bibitem[Sch{\"a}fer et~al.(2020)Sch{\"a}fer, Schlecht, and Rabenstein]{schafer2020string}
Maximilian Sch{\"a}fer, Sebastian Schlecht, and Rudolf Rabenstein.
\newblock A string in a room: Mixed-dimensional transfer function models for sound synthesis.
\newblock In \emph{International Conference on Digital Audio Effects}, pages 187--194. DAFx, 2020.

\end{thebibliography}


\newpage
\appendix

\section{Proof of the Linear Damped Stiff String Solution}\label{appendix: linear string vibration}

\begin{proposition}
A solution to the damped linear stiff string model \autoref{eqn:linear-wave}
with a clamped boundary condition $u(\pm L/2,t)=u_{x}(\pm L/2,t)=0$
and the initial condition given as $u(x,0)=u_0(x)$ and $\partial_t u(x,0)=0$
can be expressed as 
\[\begin{split}
    u(x,t) &= \sum_{n=1}^\infty X_n(x) T_n(t) \\
    X_n(x) &= c_1\left(\sin\mu_n x - \frac{\sin\mu_n L/2}{\sinh\nu_n L/2}\sinh\nu_n x\right)
            + c_2\left(\cos\mu_n x - \frac{\cos\mu_n L/2}{\cosh\nu_n L/2}\cosh\nu_n x\right) \\
    T_n(t) &= e^{-\sigma_0 t} \cos\left(\sqrt{\mu_n^4\kappa^2 + \mu_n^2 \gamma^2 - \sigma_0^2}\right) t
\end{split}\]
\end{proposition}
\begin{proof}
As it is hinted in \autoref{eqn: analytic-u},
the procedure to derive \autoref{eqn: analytic-X} and \autoref{eqn: analytic-T}
uses the method of separation of variables.
The derivation consists of three main steps that start by trying the ansatz
\(u(x,t) = X(x)T(t)\).
Substituting this ansatz into the \autoref{eqn:linear-wave} gives
\begin{equation}\label{eqn: modal solution three cases}
    \gamma^2\frac{X''}{X} - \kappa^2\frac{X^{(4)}}{X} = \frac{T''}{T} + 2\sigma_0\frac{T'}{T} = \begin{cases}
        \varsigma \\
        0 \\
        -\varsigma \\
    \end{cases}       
\end{equation}
with $\varsigma\in\mathbb{R}$ as the separation constant.
\begin{enumerate}

    \item \textbf{Solving for \(T\).}
    \begin{equation}
        T'' + 2\sigma_0 T \pm\varsigma^2 T = 0
    \end{equation}
    Roots of the characteristic polynomial of this equation are \(\beta_\pm=-\sigma_0\pm\sqrt{\sigma_0^2\mp\varsigma^2}\).
    Three solutions are available:
    \begin{itemize}
        \item Overdamping ($\sigma_0^2 > \varsigma^2$): $T = A_1 e^{\beta_+ t} + A_2 e^{\beta_- t}$
    
        \item Critical damping ($\sigma_0^2 = \varsigma^2$): $T = (A_1 + A_2 t) e^{-\sigma_0 t}$
        \item Underdamping ($\sigma_0^2 < \varsigma^2$): Rewrite the roots as $\beta_+ = -\sigma_0 + i\omega$ and $\beta_- = -\sigma_0 - \hat{\omega}$ with $\omega=\sqrt{\varsigma^2-\sigma_0^2}, \hat{\omega}=\sqrt{\varsigma^2+\sigma_0^2}$, then
        \begin{equation}\label{eqn: solution T}
            T = e^{-\sigma_0 t} (A_1 e^{i\omega t} + A_2 e^{-\hat{\omega} t})
        \end{equation} 
        The initial condition $\partial_t u(x,0)=0$ implies $A_2=0$ yielding the real solution of the form $T=A_1 e^{-\sigma_0 t} \cos\omega t$. 
        This implies that the only valid root is $\beta_+=-\sigma_0+i\omega$, hence it is enough to consider the $-\varsigma$ case only, in the \autoref{eqn: modal solution three cases}.
    \end{itemize}
    This work only considers the underdamped case where the $\sigma_0$ is sufficiently small,
    as it should be in a reasonable string model for any musical purposes.

    \item \textbf{Solving for \(X\).}
    \begin{equation}
        X^{(4)} - \frac{\gamma^2}{\kappa^2} X'' -\frac{\varsigma^2}{\kappa^2} X = 0
    \end{equation}
    Substitute \(l=\gamma^2/2\kappa^2\) and \(m=\varsigma^2/\kappa^2\).
    The roots of the characteristic polynomial for this equation are \(\pm\sqrt{l\pm\sqrt{l^2+m}}\) where \(l-\sqrt{l^2+m}\leq0\).
    Therefore the solution is rewritten as
    \begin{equation}
        X(x) = \underbrace{B_1 \sin\mu x + B_2 \sinh\nu x}_{\text{odd function}} + \underbrace{B_3 \cos\mu x + B_4 \cosh\nu x}_{\text{even function}}
    \end{equation}
    where \(\mu=\sqrt{\sqrt{l^2+m}-l}\) and \(\nu=\sqrt{\sqrt{l^2+m}+l}\).
    By applying the boundary condition \(u(L/2, t)=0\) to the even and odd functions each, then obtain
    \begin{equation}
        B_2 = -\frac{\sin\mu\frac{L}{2}}{\sinh\nu\frac{L}{2}}B_1
        \quad\text{and}\quad
        B_4 = -\frac{\cos\mu\frac{L}{2}}{\cosh\nu\frac{L}{2}}B_3
    \end{equation}
    reducing the solution as
    \begin{equation}\label{eqn: solution X}
        X(x) = B_1 \left(\sin\mu x -\frac{\sin\mu\frac{L}{2}}{\sinh\nu\frac{L}{2}} \sinh\nu x\right) + B_3 \left(\cos\mu x -\frac{\cos\mu\frac{L}{2}}{\cosh\nu\frac{L}{2}} \cosh\nu x\right).
    \end{equation}

    \item \textbf{Finding allowed values that satisfy the conditions.}\\
    Substituting the ansatz by \autoref{eqn: solution T} and \autoref{eqn: solution X} summarizes the solution as
    \begin{equation}
    \begin{split}
        u(x,t) &= e^{-\sigma_0 t}\cos\left(\sqrt{\mu^4\kappa^2+\mu^2\gamma^2 - \sigma_0^2}\cdot t\right) \\
               &\times  \Bigg[ c_1 \left(\sin\mu x -\frac{\sin\mu\frac{L}{2}}{\sinh\nu\frac{L}{2}} \sinh\nu x\right) + c_2 \left(\cos\mu x -\frac{\cos\mu\frac{L}{2}}{\cosh\nu\frac{L}{2}} \cosh\nu x\right)\Bigg] 
    \end{split}
    \end{equation}
    As mentioned earlier, the coefficients \(c_1=A_1B_1\) and \(c_2=A_1B_3\)
    are determined by the allowed values that satisfy the initial condition
    \(\sum_{n=1}^\infty X_n(x)=u_0(x)\).
    Yet, this is usually preceded by obtaining the $\mu$ and $\nu$ values.
    To determine the allowed values of $\mu$,
    apply the boundary condition $u(L/2, t)=u_x(L/2, t)=0$ to the even function that gives
    \begin{equation}
        B_3\cos\mu\frac{L}{2} = -B_4\cosh\nu\frac{L}{2}
        \quad\text{and}\quad
        -\mu B_3\sin\mu\frac{L}{2} = -\nu B_4\sinh\nu\frac{L}{2}
    \end{equation}
    or alternatively,
    \begin{equation}\label{eqn: mu tan}
        \mu\tan\mu\frac{L}{2} = -\nu\tanh\nu\frac{L}{2}.
    \end{equation}
    Similarly, applying the same boundary condition to the odd function gives
    \begin{equation}
        B_3\sin\mu\frac{L}{2} = -B_4\sinh\nu\frac{L}{2}
        \quad\text{and}\quad
        \mu B_3\cos\mu\frac{L}{2} = -\nu B_4\cosh\nu\frac{L}{2}
    \end{equation}
    or equivalently,
    \begin{equation}\label{eqn: nu tan}
        \nu\tan\mu\frac{L}{2} = \mu\tanh\nu\frac{L}{2}.
    \end{equation}
    By finding the values of \(\mu\) and \(\nu=\sqrt{\mu^2+2l}\)
    that satisfy the \autoref{eqn: mu tan} and \autoref{eqn: nu tan}
    as \(\mu_n\) and \(\nu_n\),
    one can determine the allowed values of
    \(m_n=(\mu_n^2+l)^2-l^2\) and \(\varsigma_n=\sqrt{m_n\kappa^2}\).
\end{enumerate}
Following these three steps gives the \(n\)-th mode oscillation of
\(u(x,t)\) as \(X_n(x)T_n(t)\) where
\begin{equation}\label{eqn: mode-X}
        X_n(x) = c_1\left(\sin\mu_n x - \frac{\sin\mu_n L/2}{\sinh\nu_n L/2}\sinh\nu_n x\right)
               + c_2\left(\cos\mu_n x - \frac{\cos\mu_n L/2}{\cosh\nu_n L/2}\cosh\nu_n x\right)
\end{equation}
and
\begin{equation}\label{eqn: mode-T}
    T_n(t) = e^{-\sigma_0 t} \cos\left(\sqrt{\mu_n^4\kappa^2 + \mu_n^2 \gamma^2 - \sigma_0^2}\right) t.
\end{equation}
Hence, the modal solution is expressed as a superposition of modes,
where the infinite summation as in \autoref{eqn: analytic-u}
finalizes the modal expression of the modal solution.
\end{proof}

\section{Nonlinear Damped Stiff String Vibration}\label{appendix: planar string vibration}
The Kirchhoff–Carrier model stands out as one of the simplest representations of nonlinear distributed strings, as it effectively captures the pitch glide effect and can be easily simulated using relatively straightforward finite difference schemes.
However, it is also known that only the transverse motion is explicitly accounted for, so the longitudinal motion is averaged across the length of the string.
Obviously, there are reports that this approximation may suffice under specific conditions \cite{bank2006physics}.
On the other hand, it is also reported that the interaction between longitudinal and transverse motion can result in generating rich timbres with perceptually crucial effects such as phantom partials \cite{bilbao2023models}.
A comprehensive model of string vibration, which incorporates both longitudinal and transverse motion within a single plane (in dimensional form), is outlined as follows:
\begin{subequations}\label{eqn: planar string dimensional}
    \begin{align}
        \rho A \partial_{tt} u &= EA \partial_{xx} u - (EA-T_0) \partial_x\left(\frac{\partial\Phi}{\partial (\partial_x u)}\right) \\
        \rho A \partial_{tt} \zeta &= EA \partial_{xx}\zeta - (EA-T_0)\partial_x \left(\frac{\partial\Phi}{\partial (\partial_x \zeta)}\right)
    \end{align}
\end{subequations}
The constants $\rho$ and $T_0$ are the material density and tension, respectively.
In order to model the stiffness, Young’s modulus $E$ and the cross-sectional area $A$ are introduced.

The function $\Phi$, which nonlinearly connects the two equations, is defined as follows:
\begin{equation}
    \Phi = \sqrt{(1+\partial_x\zeta)^2 + \partial_x u^2} - 1 - \partial_x \zeta
\end{equation}
Note that the final term $-1 - \partial_x\zeta$ does not affect the
dynamics of system \autoref{eqn: planar string dimensional};
it is included solely to adjust the zero-point energy of the entire system.
Substituting $x' = x/L$, $u' = u/L$, and $\zeta' = \zeta/L$ into the above system,
and then removing the primes, one obtains:
\begin{subequations}\label{eqn: string continuous phi}
    \begin{align}
        \partial_{tt} u &= \gamma^2 \partial_{xx} u - \gamma^2(\alpha^2-1)\partial_x \left(\frac{\partial\Phi}{\partial q}\right) \\
        \partial_{tt}\zeta &= \gamma^2\alpha^2 \partial_{xx}\zeta - \gamma^2(\alpha^2-1)\partial_x \left(\frac{\partial\Phi}{\partial p}\right)
    \end{align}
\end{subequations}
where the parameters $\gamma$ and $\alpha$ are defined by
\begin{equation}
    \gamma = \frac{1}{L}\sqrt{\frac{T_0}{\rho A}} \qquad\text{and}\qquad \alpha=\sqrt{\frac{EA}{T_0}}.
\end{equation}
For $\Phi$, one has \(\Phi (p,q) = \sqrt{(1+p)^2 + q^2} - 1 - p\).
Series approximations have been instrumental in analyzing nonlinear systems
like the string model mentioned earlier;
approximations up to third or fourth order are frequently used \cite{morse1986theoretical}.
The function $\Phi(p,q)$ can be approximated with a variety of orders in $p$ and $q$:
\[\Phi_2=\frac{1}{2}q^2\qquad \Phi_3=\frac{1}{2}q^2 - \frac{1}{2}pq^2\qquad \Phi_4=\frac{1}{2}q^2 - \frac{1}{2}pq^2 + \frac{1}{2}q^2p^2-\frac{1}{8}q^4\]
In this paper, we consider the following approximation to $\Phi(p,q)$, similar to $\Phi_4$, but lacking one of the fourth-order terms, following \citet{bilbao2009numerical}.
\[\Phi_4^* = \frac{1}{2}q^2-\frac{1}{2}pq^2 - \frac{1}{8}q^4\]
Under this choice $\Phi_4^*$, the system \autoref{eqn: string continuous phi} reduces to
\begin{subequations}\label{eqn: string continuous series approx}
    \begin{align}
        \partial_{tt} u &= \gamma^2\alpha^2 \partial_{xx} u - \gamma^2(\alpha^2-1)\partial_x\left(q^3+2pq\right) \\
        \partial_{tt} \zeta &= \gamma^2\alpha^2 \partial_{xx}\zeta - \gamma^2(\alpha^2-1)\partial_x\left(q^2\right)
    \end{align}
\end{subequations}
which can be augmented by the stiffness and the damping term to become \autoref{eqn: wave-nonlinear}.
The nonlinear damped stiff string system as \autoref{eqn: wave-nonlinear} is very similar to the work by \citet{bank2004piano} but with the damping coefficients given as the same values for the longitudinal and transverse directions.

It is also noteworthy that there are many approaches to modeling the nonlinear string vibration, other than the aforementioned Kirchhoff--Carrier-style one.
A good example is the Functional Transformation Method (FTM), which is a powerful method for modeling the oscillation of acoustic systems in terms of transfer functions.
Interested readers are advised to read its derivation and applications from \cite{trautmann2012digital,schafer2016physical,rabenstein2009tubular,schafer2020string}.
Those with a background in neural network frameworks may also find reading \cite{schlecht2022physical} particularly engaging, as \citet{schlecht2022physical} points out the connection between the underlying mathematical ideas in the FTMs and the FNOs.

\section{Damping coefficient}\label{appendix: damping coefficients}
The damping coefficients \(\sigma_0\) and \(\sigma_1\) are typically determined
experimentally, especially those for nonlinear systems.
However, it can be somewhat difficult to set the values without any prior knowledge of the observations.
In this regard, the practical settings for the damping coefficients
can be approximately derived from the decay times, by resorting to the values for linear systems.
The linear string damping coefficients \(\sigma_0\) and \(\sigma_1\) are derived from the (frequency-dependent) T60 values:
where $f_{\mathrm{T60}}^{(1)}$ and $f_{\mathrm{T60}}^{(2)}$ denotes the two distinct frequencies,
and $t_{\mathrm{T60}}^{(1)}$ and $t_{\mathrm{T60}}^{(2)}$ denotes the corresponding decay times.
In this case,
\begin{equation}\label{eqn: t60}
    \sigma_0 = \frac{6\log(10)}{\xi_1-\xi_2} \left(\frac{\xi_1}{t_{\mathrm{T60}}^{(2)}} - \frac{\xi_2}{t_{\mathrm{T60}}^{(1)}}\right),
    \qquad
    \sigma_1 = \frac{6\log(10)}{\xi_1-\xi_2} \left(\frac{1}{t_{\mathrm{T60}}^{(1)}} - \frac{1}{t_{\mathrm{T60}}^{(2)}}\right),
\end{equation}
where
\(\xi_i = - \gamma^2 + \sqrt{\gamma^4 + 4\kappa^2 \times (2 \pi f_{\mathrm{T60}}^{(i)})^2}\)
for \(i=1,2\).
For this model, as the energy loss increases monotonically with frequency,
one must choose $t_{\mathrm{T60}}^{(1)}\geq t_{\mathrm{T60}}^{(2)}$ when $f_{\mathrm{T60}}^{(1)} < f_{\mathrm{T60}}^{(2)}$.

\begin{figure}
    \centering
    \centerline{\includegraphics[width=0.6\linewidth]{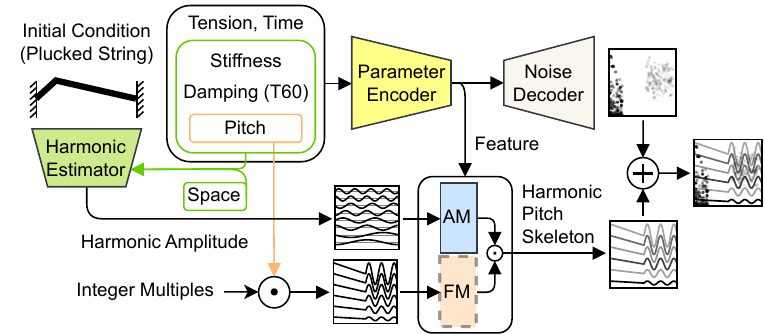}}
    \caption{
        The DDSPish is designed in a similar way to DMSP,
    }               
    \label{fig: ddsp architecture}
\end{figure}

\section{Baselines}\label{appendix: baseline}
In this section, we present more details of the baselines.
All models, including DMSP, are trained using RAdam optimizer, with Noam learning rate scheduler with a peak learning rate of $10^{-3}$ reaching at 1000 number of warmup steps.
\autoref{fig: ddsp architecture} summarizes the architectures of DDSPish models.
The only architectural difference between DDSPish and DDSPish-{\scriptsize XFM} is the existence of the FM module.
DDSPish-{\scriptsize XFM} is \textit{harmonic} as the mode frequencies are not modulated after its initialization by the integer multiples of $f_0$, while DDSPish is \textit{inharmonic} as FM is trained to modulate the harmonic pitch skeleton to match the inharmonic mode frequencies.

For the Modal synthesis,
we compute the solution \autoref{eqn:linear-wave}
with the allowed values of \(\mu_n\), \(\nu_n\), \(c_1\), and \(c_2\)
obtained using the Levenberg--Marquardt algorithm.
We compute the modes up to the 100\textsuperscript{th} order
with the double-precision floating-point arithmetics.
The modes are then post-processed to be cut under the Nyquist limit.
Subsequently, all models are trained using 40 numbers of modes.
We use 65 number of bands used for the filtered noise.

\begin{table}[h]
  \caption{PDE parameter sampling range}
  \label{tab: params}
  \centering
  \begin{tabular}{l rrr}
    \toprule
    & Min. & Max. & Unit \\
    \midrule
    $f_0$ & 98.00 & 440.0 & Hz \\
    $\kappa$ & $0.01\gamma$ & $0.03\gamma$ & - \\
    $\alpha$ & 1 & 25 & - \\
    $t_{\mathrm{T60}}^{(1)}$ & 10 & 25 & sec \\
    $t_{\mathrm{T60}}^{(2)}$ & 10 & 30 & sec \\
    $f_{\mathrm{T60}}^{(1)}$ & 1100 & 1200 & Hz \\
    $f_{\mathrm{T60}}^{(2)}$ & 100 & $f_{\mathrm{T60}}^{(1)}-1000$ & Hz \\
    \midrule
    $p_a$ & 0.001 & 0.02 & - \\
    $p_x$ & 0.1 & 0.5 & - \\
    \bottomrule
  \end{tabular}
  \vspace{-5mm}
\end{table}

\section{Datasets}\label{appendix: datasets}
As described in \autoref{ssec: experimental setup}, this paper utilizes the nonlinear string simulator, StringFDTD-Torch, presented by \citet{lee2024string}.
For the simulation, PDE parameters are uniformly random-sampled within the moderate parameter ranges.
\autoref{tab: params} summarizes the infimum and the supremum values for each uniform distribution of the PDE parameter.
For a given random-sampled parameter set, the simulator outputs the transverse and the longitudinal solutions ($u$ and $\zeta$, resp.) while this work considers $u$ only.
Yet, it is worth mentioning that the adopted $u$ is different from that of the one-dimensional string since $u$ and $\zeta$ are \textit{coupled} as evident in \autoref{eqn: wave-nonlinear}.
The obtained $u$ is defined over a spatio-temporal grid, where the spatial resolutions are carefully chosen to mitigate the numerical stability criteria and the numerical dispersions while keeping the temporal resolution by the prefixed audio sampling rate.
As the grid spacing is fixed by these values, the raw simulation data are consisted of diverse grid sizes depending on $f_0$, $\kappa$, and $\alpha$, making it difficult to batchify for training.
For this reason, we spatially upsample the data to a fixed spatial grid size of 256, using a bivariate spline approximation over a rectangular spatio-temporal mesh up to the 5\textsuperscript{th} order degree.



\newpage
\section*{NeurIPS Paper Checklist}

\begin{enumerate}

\item {\bf Claims}
    \item[] Question: Do the main claims made in the abstract and introduction accurately reflect the paper's contributions and scope?
    \item[] Answer: \answerYes{} 
    \item[] Justification: We reflect the paper's contributions and scope to the main claims.

\item {\bf Limitations}
    \item[] Question: Does the paper discuss the limitations of the work performed by the authors?
    \item[] Answer: \answerYes{} 
    \item[] Justification: We discuss the limitations in the conclusion section. 

\item {\bf Theory Assumptions and Proofs}
    \item[] Question: For each theoretical result, does the paper provide the full set of assumptions and a complete (and correct) proof?
    \item[] Answer: \answerYes{} 
    \item[] Justification: We included the assumptions and proofs in the manuscript.

    \item {\bf Experimental Result Reproducibility}
    \item[] Question: Does the paper fully disclose all the information needed to reproduce the main experimental results of the paper to the extent that it affects the main claims and/or conclusions of the paper (regardless of whether the code and data are provided or not)?
    \item[] Answer: \answerYes{} 
    \item[] Justification: We tried our best to fully disclose all the information needed to reproduce the results, and are willing to supplement any additional information, if needed.

\item {\bf Open access to data and code}
    \item[] Question: Does the paper provide open access to the data and code, with sufficient instructions to faithfully reproduce the main experimental results, as described in supplemental material?
    \item[] Answer: \answerYes{} 
    \item[] Justification: We provide code and data attached in the link at the footnote.

\item {\bf Experimental Setting/Details}
    \item[] Question: Does the paper specify all the training and test details (e.g., data splits, hyperparameters, how they were chosen, type of optimizer, etc.) necessary to understand the results?
    \item[] Answer: \answerYes{} 
    \item[] Justification: We specify the training and test details.

\item {\bf Experiment Statistical Significance}
    \item[] Question: Does the paper report error bars suitably and correctly defined or other appropriate information about the statistical significance of the experiments?
    \item[] Answer: \answerYes{} 
    \item[] Justification: We report the experimental results using appropriate statistics of the measured scores.

\item {\bf Experiments Compute Resources}
    \item[] Question: For each experiment, does the paper provide sufficient information on the computer resources (type of compute workers, memory, time of execution) needed to reproduce the experiments?
    \item[] Answer: \answerYes{} 
    \item[] Justification: We provide the computational resources for the dataset construction and the computation.
    
\item {\bf Code Of Ethics}
    \item[] Question: Does the research conducted in the paper conform, in every respect, with the NeurIPS Code of Ethics \url{https://neurips.cc/public/EthicsGuidelines}?
    \item[] Answer: \answerYes{} 
    \item[] Justification: The research conducted in this paper conform with the NeurIPS code of ethics.

\item {\bf Broader Impacts}
    \item[] Question: Does the paper discuss both potential positive societal impacts and negative societal impacts of the work performed?
    \item[] Answer: \answerNA{} 
    \item[] Justification: The research proposes neural network for the physical simulation of musical instrument for scientific computing.
    
\item {\bf Safeguards}
    \item[] Question: Does the paper describe safeguards that have been put in place for responsible release of data or models that have a high risk for misuse (e.g., pretrained language models, image generators, or scraped datasets)?
    \item[] Answer: \answerNA{} 
    \item[] Justification: The paper does not poses such risks.

\item {\bf Licenses for existing assets}
    \item[] Question: Are the creators or original owners of assets (e.g., code, data, models), used in the paper, properly credited and are the license and terms of use explicitly mentioned and properly respected?
    \item[] Answer: \answerYes{} 
    \item[] Justification: We tried our best to include as many proper credits for the original owners as possible by citing the original works. We will include the missing credits if any.

\item {\bf New Assets}
    \item[] Question: Are new assets introduced in the paper well documented and is the documentation provided alongside the assets?
    \item[] Answer: \answerYes{} 
    \item[] Justification: We documented the code and its usage properly, and will maintain the documentation.

\item {\bf Crowdsourcing and Research with Human Subjects}
    \item[] Question: For crowdsourcing experiments and research with human subjects, does the paper include the full text of instructions given to participants and screenshots, if applicable, as well as details about compensation (if any)? 
    \item[] Answer: \answerNA{} 
    \item[] Justification: The paper does not involve crowdsourcing nor research with human subjects.

\item {\bf Institutional Review Board (IRB) Approvals or Equivalent for Research with Human Subjects}
    \item[] Question: Does the paper describe potential risks incurred by study participants, whether such risks were disclosed to the subjects, and whether Institutional Review Board (IRB) approvals (or an equivalent approval/review based on the requirements of your country or institution) were obtained?
    \item[] Answer: \answerNA{} 
    \item[] Justification: The paper does not involve crowdsourcing nor research with human subjects.

\end{enumerate}


\end{document}